\pdfoutput=1
\newcommand*{\ATLASLATEXPATH}{}
\documentclass[USenglish,texlive=2016,cernpreprint]{\ATLASLATEXPATH atlasdoc}
\usepackage{\ATLASLATEXPATH atlaspackage}
\usepackage{\ATLASLATEXPATH atlasbiblatex}
\usepackage{\ATLASLATEXPATH atlasphysics}
\usepackage{amssymb}
\usepackage{relsize}
\usepackage{xspace}
\xspaceaddexceptions{]\}}
\usepackage{rotating}
\usepackage{multirow}
\usepackage{multicol}
\usepackage{dcolumn}
\usepackage{listings}
\usepackage{enumitem}
\usepackage{color}
\usepackage[utf8]{inputenc}

\newcommand{\powheg} {\textsc{Powheg-Box}\xspace}
\newcommand{\pythia} {\textsc{Pythia6}\xspace}
\newcommand{\pythiapp} {\textsc{Pythia8}\xspace}
\newcommand{\prospino} {\textsc{Prospino2}\xspace}

\newcommand{\madgraph} {\textsc{MG5\_aMC@NLO}\xspace}

\newcommand{\geant} {\textsc{Geant4}\xspace}

\newcommand{\rhad} {\mbox{$R$-hadron}\xspace}
\newcommand{\rhads} {\mbox{$R$-hadrons}\xspace}

\newcommand{\lumi} {36.1~\ifb\xspace}
\newcommand{\lumiUnc}{2.1\%\xspace}

\renewcommand{\pt} {\ensuremath{p_{\mathrm{T}}}\xspace}

\newcommand{\zmumu} {\ensuremath{Z \to \mu\mu}\xspace}
\newcommand{\etmiss} {\ensuremath{E_{\mathrm{T}}^\mathrm{miss}}\xspace}
\newcommand{\ptmiss} {\ensuremath{\vec{p}_{\mathrm{T}}^\mathrm{~miss}}\xspace}
\newcommand{\CLs} {\ensuremath{\mathrm{CL}_\mathrm{s}}\xspace}
\newcommand{\pb} {\ensuremath{\mathrm{{pb}}}\xspace}
\renewcommand{\MeV} {\ensuremath{\mathrm{{Me\kern -0.1em V}}}\xspace}
\renewcommand{\GeV} {\ensuremath{\mathrm{{Ge\kern -0.1em V}}}\xspace}
\renewcommand{\TeV} {\ensuremath{\mathrm{{Te\kern -0.1em V}}}\xspace}

\newcommand{\MeVgcm} {\ensuremath{\MeV\mathrm{g}^{-1}\mathrm{cm}^{2}}\xspace}
\newcommand{\dedx} {{\ensuremath{dE/dx}}\xspace}

\newcommand{\pzero} {\ensuremath{p_0}\xspace}
\newcommand{\splitcell}[2][b]{\begin{tabular}[#1]{@{}c@{}}#2\end{tabular}}

\newcommand{\tof} {{ToF}\xspace}
\newcommand{\bgdedx} {\ensuremath{(\beta\gamma)_{dE/dx}}\xspace}

\newcommand{\betatile} {\ensuremath{\beta_\mathrm{\textsc{tile}}}\xspace}
\newcommand{\betatilehit} {\ensuremath{\beta^\mathrm{\textsc{hit}}_\mathrm{\textsc{tile}}}\xspace}
\newcommand{\betamdt} {\ensuremath{\beta_\mathrm{\textsc{mdt}}}\xspace}
\newcommand{\betarpc} {\ensuremath{\beta_\mathrm{\textsc{rpc}}}\xspace}

\newcommand{\betatof} {\ensuremath{\beta_\mathrm{\textsc{ToF}}}\xspace}
\newcommand{\msagno} {MS-agnostic\xspace}
\newcommand{\mtof} {\ensuremath{m_\mathrm{ToF}}\xspace}
\newcommand{\mdedx} {\ensuremath{m_{dE/dx}}\xspace}

\newcommand{\selidcalo} {\textsc{\texttt{id{\smaller +}calo}}\xspace}
\newcommand{\selloose} {\textsc{\texttt{loose}}\xspace}

\newcommand{\seltight} {\textsc{\texttt{tight}}\xspace}

\newcommand{\srrhma} {\texttt{SR-Rhad-MSagno}\xspace}
\newcommand{\srrhfd} {\texttt{SR-Rhad-FullDet}\xspace}
\newcommand{\srone} {\texttt{SR-1Cand-FullDet}\xspace}
\newcommand{\srtwo} {\texttt{SR-2Cand-FullDet}\xspace}

\newcommand{\mtofmin} {\ensuremath{m_\mathrm{ToF}^\text{min}}\xspace}
\newcommand{\mdedxmin} {\ensuremath{m_{dE/dx}^\text{min}}\xspace}
\newcommand{\limRHMAgluExp} {{2060~\GeV}\xspace}
\newcommand{\limRHMAgluObs} {{1950~\GeV}\xspace}
\newcommand{\limRHMAsboExp} {{1270~\GeV}\xspace}
\newcommand{\limRHMAsboObs} {{1190~\GeV}\xspace}
\newcommand{\limRHMAstoExp} {{1345~\GeV}\xspace}
\newcommand{\limRHMAstoObs} {{1265~\GeV}\xspace}
\newcommand{\limRHFDgluExp} {{2050~\GeV}\xspace}
\newcommand{\limRHFDgluObs} {{2000~\GeV}\xspace}
\newcommand{\limRHFDsboExp} {{1280~\GeV}\xspace}
\newcommand{\limRHFDsboObs} {{1250~\GeV}\xspace}
\newcommand{\limRHFDstoExp} {{1370~\GeV}\xspace}
\newcommand{\limRHFDstoObs} {{1340~\GeV}\xspace}
\newcommand{\limStauExp} {{420~\GeV}\xspace}
\newcommand{\limStauObs} {{430~\GeV}\xspace}
\newcommand{\limCharExp} {{1070~\GeV}\xspace}
\newcommand{\limCharObs} {{1090~\GeV}\xspace}
\newcommand{\limRHMAgluExpFivty} {{2090~\GeV}\xspace}
\newcommand{\limRHMAgluObsFivty} {{1980~\GeV}\xspace}
\newcommand{\limRHMAgluExpThrity} {{2080~\GeV}\xspace}
\newcommand{\limRHMAgluObsThrity} {{1960~\GeV}\xspace}
\newcommand{\limRHMAgluExpTen} {{1980~\GeV}\xspace}
\newcommand{\limRHMAgluObsTen} {{1860~\GeV}\xspace}

\makeatletter
\newcommand\footnoteref[1]{\protected@xdef\@thefnmark{\ref{#1}}\@footnotemark}
\makeatother

\AtlasTitle{\Large Search for heavy charged long-lived particles in the ATLAS detector in \lumi of proton--proton collision data at $\sqrt{s} = 13~\TeV$}

\author{The ATLAS Collaboration}
\AtlasRefCode{SUSY-2016-32}
\PreprintIdNumber{CERN-EP-2018-339}
\AtlasDate{\today}
\AtlasJournal{Phys.\ Rev.\ D}
\AtlasJournalRef{Phys. Rev. D 99 (2019) 092007}
\AtlasDOI{DOI:10.1103/PhysRevD.99.092007}
\AtlasAbstract{%
A search for heavy charged long-lived particles is performed using a data sample of \lumi of proton--proton collisions at $\sqrt{s} = 13~\TeV$ collected by the ATLAS experiment at the Large Hadron Collider. The search is based on observables related to ionization energy loss and time of flight, which are sensitive to the velocity of heavy charged particles traveling significantly slower than the speed of light. Multiple search strategies for a wide range of lifetimes, corresponding to path lengths of a few meters, are defined as model-independently as possible, by referencing several representative physics cases that yield long-lived particles within supersymmetric models, such as gluinos/squarks (\rhads), charginos and staus. No significant deviations from the expected Standard Model background are observed. Upper limits at 95\% confidence level are provided on the production cross sections of long-lived \rhads as well as directly pair-produced staus and charginos. These results translate into lower limits on the masses of long-lived gluino, sbottom and stop \rhads, as well as staus and charginos of \limRHFDgluObs, \limRHFDsboObs, \limRHFDstoObs, \limStauObs and \limCharObs, respectively.
}

\hypersetup{pdftitle={ATLAS document},pdfauthor={The ATLAS Collaboration}}

\begin{document}
\maketitle



\section{Introduction}
\label{sec:intro}

The search for heavy charged long-lived particles presented in this paper is based on a data sample of \lumi of proton--proton ($pp$) collisions at $\sqrt{s} = 13~\TeV$ collected in 2015 and 2016. It utilizes observables related to large ionization energy loss (\dedx) and time of flight (\tof), which are signatures of heavy charged particles traveling significantly slower than the speed of light. The mass of the particles is estimated using the \dedx (\mdedx) or \tof (\mtof) measurements together with the reconstructed momentum. The background is estimated in a purely data-driven manner, and multiple signal regions are defined to address the different possible signatures of heavy long-lived particles (LLPs) that reach at least the ATLAS hadronic calorimeter, which corresponds to decay lengths of a few meters. Previous searches for LLPs that are stable within the detector were performed by the CMS Collaboration using 2.5~\ifb\xspace of data at 13~\TeV~\cite{Khachatryan:2016sfv}, as well as by the ATLAS Collaboration using 3.2~\ifb\xspace of data at 13~\TeV~\cite{Aaboud:2016uth} and 19.8~\ifb\xspace of data at 8~\TeV~\cite{ATLAS:2014fka}.

LLPs are predicted in a variety of theories that extend the Standard Model (SM)~\cite{Fairbairn:2006gg}. Theories with supersymmetry (SUSY)~\cite{Golfand:1971iw, Volkov:1973ix, Wess:1974tw, Wess:1974jb, Ferrara:1974pu, Salam:1974ig}, which either violate~\cite{Dreiner:1997uz, Berger:2003kc, Barbier:2004ez} or conserve~\cite{Farrar:1978xj, Fairbairn:2006gg, Kolda:1997wt, Baer:1998pg, Gates:1999ei, Giudice:2004tc, ArkaniHamed:2004yi} $R$-parity, allow for the existence of charged long-lived sleptons ($\tilde{\ell}$), squarks ($\tilde{q}$), gluinos ($\tilde{g}$) and charginos ($\tilde{\chi}^{\pm}_{1}$).

Colored LLPs (e.g. $\tilde{q}$ and $\tilde{g}$) would hadronize forming so-called \rhads~\cite{Farrar:1978xj}, which are bound states composed of the LLP and light SM quarks or gluons, and may emerge from the collision as charged or neutral states. Through hadronic interactions of the light-quark constituents with the detector material, especially inside the calorimeters, \rhads can change to states with a different electric charge. Thus they might not be reconstructed as a consistently charged track in the inner tracking detector (ID) and in the muon spectrometer (MS), even if the lifetime is long enough to traverse the entire detector. Searches for \rhads are performed following two different approaches: using all available detector information (``full-detector \rhad search''), or disregarding all information from the muon spectrometer (``\msagno \rhad search'') to minimize the dependence on the modeling of \rhad interactions with the material of the detector. Long-lived gluinos are motivated for example by split-SUSY models~\cite{Giudice:2004tc, ArkaniHamed:2004yi}, in which high-mass squarks can lead to very long gluino lifetimes. Long-lived squarks, in particular a light top squark (stop) as the next-to-lighest SUSY particle, is motivated for example by electroweak baryogenesis~\cite{Carena:2002wz, Balazs:2004bu}, where non-universal squark mass terms can lead to a small mass difference between the stop and the neutralino as the lightest supersymmetric particle (LSP), and the lightest chargino is heavy, leading to suppressed radiative decays and long stop lifetimes.

If the charged LLP does not interact hadronically, it would predominantly lose energy via ionization as it passes through the ATLAS detector. Searches for long-lived charginos and sleptons (focusing on staus, as they are expected in most models to be the lightest) identified in both the ID and MS are therefore performed. The searches for staus are motivated by gauge-mediated SUSY breaking (GMSB)~\cite{Dine:1981gu, AlvarezGaume:1981wy, Nappi:1982hm, Dine:1993yw, Dine:1994vc, Dine:1995ag, Kolda:1997wt} assuming the LSP to be a gravitino and with the light stau ($\tilde{\tau}$) as an LLP, decaying to a $\tau$-lepton and a gravitino with an unconstrained lifetime. While essentially all events in the GMSB models that include long-lived staus involve cascade decays that end in two LLPs, in this paper direct di-stau production through a Drell--Yan process is taken as a benchmark. Results for pair-produced charginos are motivated by a minimal anomaly-mediated supersymmetry breaking (mAMSB) model, where often the supersymmetric partners of the SM $W$-boson fields, the wino fermions, are the lightest gaugino states. In this particular case, the lightest of the charged mass eigenstates, a chargino, and the lightest of the neutral mass eigenstates, a neutralino, are both almost pure wino and nearly mass-degenerate, resulting in long-lived charginos (see Refs.~\cite{Randall:1998uk, Giudice:1998xp} for details).

This paper is organized as follows. A brief description of the ATLAS detector is given in Section~\ref{sec:detector} with emphasis on the parts relevant for this analysis, followed by details of the calibration of key observables in Section~\ref{sec:obsandcal}. The dataset and simulated event samples and the subsequent event selection are described in Sections~\ref{sec:samples} and \ref{sec:selection}, respectively. Section~\ref{sec:background} explains the method of background estimation and compares these estimates with data. Section~\ref{sec:systematics} details the origin and estimation of systematic uncertainties. The results including upper cross-section limits are shown in Section~\ref{sec:result}, and finally the conclusions are summarized in Section~\ref{sec:conclusion}.

\section{ATLAS detector}
\label{sec:detector}

The ATLAS detector~\cite{PERF-2007-01} is a multipurpose particle detector consisting of the ID immersed in a 2~T solenoidal magnetic field, electromagnetic as well as hadronic calorimeters and a MS based on three large air-core toroid superconducting magnets with eight coils each. The ID comprises a silicon pixel detector, a silicon microstrip detector (SCT) and a transition-radiation tracker. With almost $4\pi$ coverage in solid angle,\footnote{ATLAS uses a right-handed coordinate system with its origin at the nominal interaction point (IP) in the center of the detector and the $z$-axis coinciding with the axis of the beam pipe. The $x$-axis points from the IP to the center of the LHC ring, and the $y$-axis points upward. Cylindrical coordinates ($r$, $\phi$) are used in the transverse plane, $\phi$ being the azimuthal angle around the beam pipe. The pseudorapidity is defined in terms of the polar angle $\theta$ as $\eta = -\ln{\tan(\theta/2)}$. Object distances in the $\eta$--$\phi$ plane are given by $\Delta R = \sqrt{\left( \Delta \eta \right)^2 + \left(\Delta \phi \right)^2}$.} the ATLAS detector is sensitive to the missing transverse momentum associated with each event. Several components are used to determine either \dedx or \tof in this search and are discussed in more detail below.

The innermost component of the ATLAS detector is the pixel detector, consisting of four radial layers of pixel-sensors in the barrel region and three disks on each side in the endcap region. The pixel detector measures the ionization energy loss of charged particles traversing it via a time-over-threshold (ToT) technique~\cite{SUSY-2016-31}.

The ATLAS calorimeter in the central detector region consists of a liquid-argon electromagnetic calorimeter followed by a steel-absorber scintillator-tile sampling calorimeter. The latter serves as hadronic calorimeter covering the region up to $|\eta|=1.7$. In the $\phi$-direction the tile calorimeter is segmented into 64 wedges with a size of $\Delta\phi = 2\pi/64 \approx 0.1$. In the $r$--$z$ plane it is formed by three radial layers in the central barrel region and three in an extended barrel on each side, where each layer is further segmented into cells. The cells in the first and third layer in the barrel and in all layers in the extended barrel have a rectangular shape, while in the second layer of the barrel they are composed of two shifted rectangles (see Figure~\ref{fig:Calib_TileCalBetaCells}). Overall, the tile calorimeter consists of 73 cells in 64 $\phi$-segments resulting in 4672 individual cells. The Optimal Filtering Algorithm (OFA)~\cite{Fullana:2005dwa} used for the readout provides, besides a precise measurement of the energy deposits of particles and jets, a timing measurement. The resolution of the single-cell timing in the tile calorimeter is 1.3--1.7~ns.

The outermost part of the ATLAS detector is the MS. Immersed in toroidal magnetic fields, the MS provides particle tracking and momentum reconstruction for charged particles with $|\eta|<2.7$, as well as triggering information in the range $|\eta|<2.4$. Three layers of muon detectors are arranged in concentric shells at distances between 5~m and 10~m from the interaction point (IP) in the barrel region, and in wheels perpendicular to the beam axis at distances between 7.4~m and 21.5~m in the endcap regions where $|\eta|>1.05$. Four different detector technologies are used in the MS. In the barrel region the muon trigger relies on resistive-plate chambers (RPCs), while thin-gap chambers are employed in the endcap wheels. High-precision tracking for the momentum measurements is performed by monitored drift tubes (MDTs), except for the innermost layer in the forward region of $2.0<|\eta|<2.7$, where cathode-strip chambers are installed. A muon traversing the detector in the barrel region typically has around 20 hits in the MDTs and 14 hits in the RPCs. Both systems exhibit a sufficiently precise single-hit timing resolution (MDTs 3.2~ns, RPCs 1.8~ns) to distinguish relativistic muons traveling at almost the speed of light from slowly propagating stable massive particles. For the reconstruction of slow particles traversing the full detector, a dedicated tracking algorithm that treats the velocity $\beta$ as a free parameter is used~\cite{Tarem:2009zz, Tarem:2010zz}.

The ATLAS trigger system~\cite{Aaboud:2016leb} consists of a hardware-based level-1 trigger followed by a software-based high-level trigger, which runs reconstruction and calibration software similar to the offline reconstruction, and reduces the event rate to about 1~kHz.

\section{Calibration of main observables}
\label{sec:obsandcal}
To achieve optimal identification performance for heavy, charged, LLPs, measurements of the specific ionization energy loss and \tof have to be calibrated.

The pixel detector provides measurements of specific ionization energy losses. The \dedx is corrected for $\eta$-dependence and on a run-by-run basis, to reduce the effect of the degradation of charge collection in the silicon sensor due to accumulated irradiation. Individual measurements resulting from a particle traversing the detector are combined in a truncated mean to reduce the effects of the Landau tails on the estimate of the most probable value (MPV). Finally, a value of the $\beta\gamma$ of the particle is estimated from the MPV and the momentum, using a three-parameter empirical function, calibrated using protons, kaons and pions with low transverse momentum (\pt). The expected resolution for $\beta\gamma$ from the \dedx and momentum measurement is about 14\%, which is taken as the uncertainty in the measured values. A detailed description of the $\beta\gamma$ estimation using the pixel detector can be found in Ref.~\cite{SUSY-2016-31}.

The particle velocity, $\beta$, is determined via \tof measurements in the tile calorimeter, the MDTs and the RPCs. To ensure an optimal $\beta$ resolution, a series of custom calibrations is performed. High-\pt muons are used for the calibration as, given the timing resolutions, they travel effectively at the speed of light.

For the tile calorimeter, only cells with a minimum energy deposit of 500~MeV are used. The time difference ($t_{0}$) relative to a particle traveling at the speed of light from the IP to the cell center (cell distance) is not allowed to be larger than 25~ns to reduce the effect of out-of-time pileup. As the first calibration step, a bias introduced by the OFA is corrected. The signal simulation reproduces the OFA bias reliably. The correction is very small for an in-time signal, while for late-arriving particles it ranges up to 10~ns. This is followed by a correction for the particle path in the cell. An effective spatial position corresponding to the timing measurement is estimated using an extrapolation of the particle track from the production vertex to the respective cell in the tile calorimeter (effective distance). The difference between the effective distance and the cell distance is used as a correction for the timing calibrations, while the effective distance is used for the estimation of the individual measurements of the velocity (\betatilehit). The corresponding correction is largest (3~ns) for the edge $\eta$ regions of the largest cells in the tile calorimeter, which are located in the outermost layer of the extended barrel. This is followed by separate $\eta$-dependent corrections for data and simulation, which accounts for small remaining $\eta$-dependences of the timing for the outer parts of the cells in data, and ensures an $\eta$-independent timing measurement in simulation. Furthermore, a calibration of the timing as a function of the energy deposit per $\phi$-projected cell (averaging cells of identical geometry over $\phi$) is applied, which does not exceed $0.2$~ns. Additionally, calibration constants in data are estimated for each cell followed by a run-by-run correction of the overall timing. These calibrations account for differences between single-tile calorimeter cells and for misalignments of the ATLAS and LHC clocks. The cell-wise calibration factors are mostly well below 1~ns, while for some cells they range up to 2~ns. The run-by-run correction constants are between $-0.6$~ns and 0.2~ns. A smearing of the timing obtained in simulation is applied to achieve the same resolution as observed in data. The width of a Gaussian parameterization of the $t_{0}$ distribution serves as the uncertainty. The uncertainties are estimated as a function of the energy deposit per $\phi$-projected cell. Finally, the uncertainties are adjusted to achieve a unit Gaussian width for the pull distribution, which gives a correction of 1\% to the uncertainty. The individual measurements of \betatilehit, with the $\phi$-projected resolutions shown in Figure~\ref{fig:Calib_TileCalBetaCells}, are combined in a weighted average \betatile using the inverse squared uncertainties as weights. The final resolution achieved for \betatile is $\sigma_{\betatile}=0.068$.\footnote{\label{sigbetafit}The resolutions ($\sigma_\text{x}$) are determined by performing a Gaussian fit to the core of the respective distributions.}

\begin{figure}[tbp]
  \centering
  \includegraphics[width=.99\textwidth]{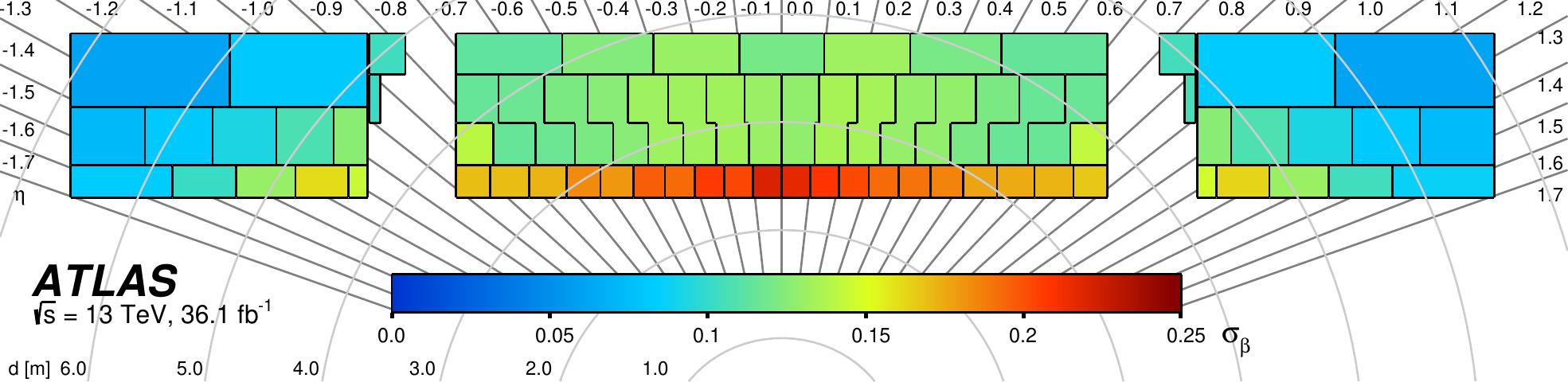}
  \caption{Resolution ($\sigma_\beta$) of the \betatilehit measurement for the different tile-calorimeter cells, which are shown with their actual shape. The gray circles indicate, the distance, $d$, from the interaction point and the gray straight lines the direction in $\eta$.}
  \label{fig:Calib_TileCalBetaCells}
\end{figure}

For the MDTs and RPCs, each of the $323\,799$ drift tubes and $362\,262$ RPC readout strips is calibrated individually by performing a Gaussian parameterization of the timing information, and correcting for the offset to the expected value for a particle traveling at the speed of light. The width of the fitted distribution is taken as the uncertainty for the measurements in the respective tube/strip. The uncertainties are, in a manner similar to the tile calorimeter, adjusted to give a unit Gaussian width for the pull distributions. The corrections to the uncertainties are 16\% for the MDTs, less than 1\% for the RPC $\eta$-strips and 2\% for the $\phi$-strips. The squared inverse of the uncertainty is used as a weight for the calculation of a weighted-average $\beta$. Time-dependent phase-shift variations between the ATLAS and LHC clocks are addressed similarly to the tile calorimeter in a separate correction of timing information for each LHC run, and separately for MDTs, RPC readout strips measuring $\eta$ and RPC strips measuring $\phi$. The calibration results in a $\beta$ resolution of $\sigma_{\betamdt}=0.026$ for the MDTs and $\sigma_{\betarpc}=0.022$ for the RPCs (with single-hit $t_0$ resolutions of about 3.2~ns and 1.8~ns, respectively).

Finally, the \tof-based $\beta$ measurements in the different subsystems are combined into an overall \betatof, which is estimated as a weighted average of the $\beta$ measurements from the different subsystems using the inverse squared uncertainties as weights. Furthermore, the combined uncertainty and the probability of compatibility between the measurements from the different subsystems are calculated for each candidate. The final distributions of the $\beta$ measurements are shown in Figure~\ref{fig:finalBeta} for muons from \zmumu selected events in data and simulation. The final resolution achieved for \betatof is $\sigma_{\betatof}=0.021$.

\begin{figure}[tbp]
  \centering
  \includegraphics[width=.49\textwidth]{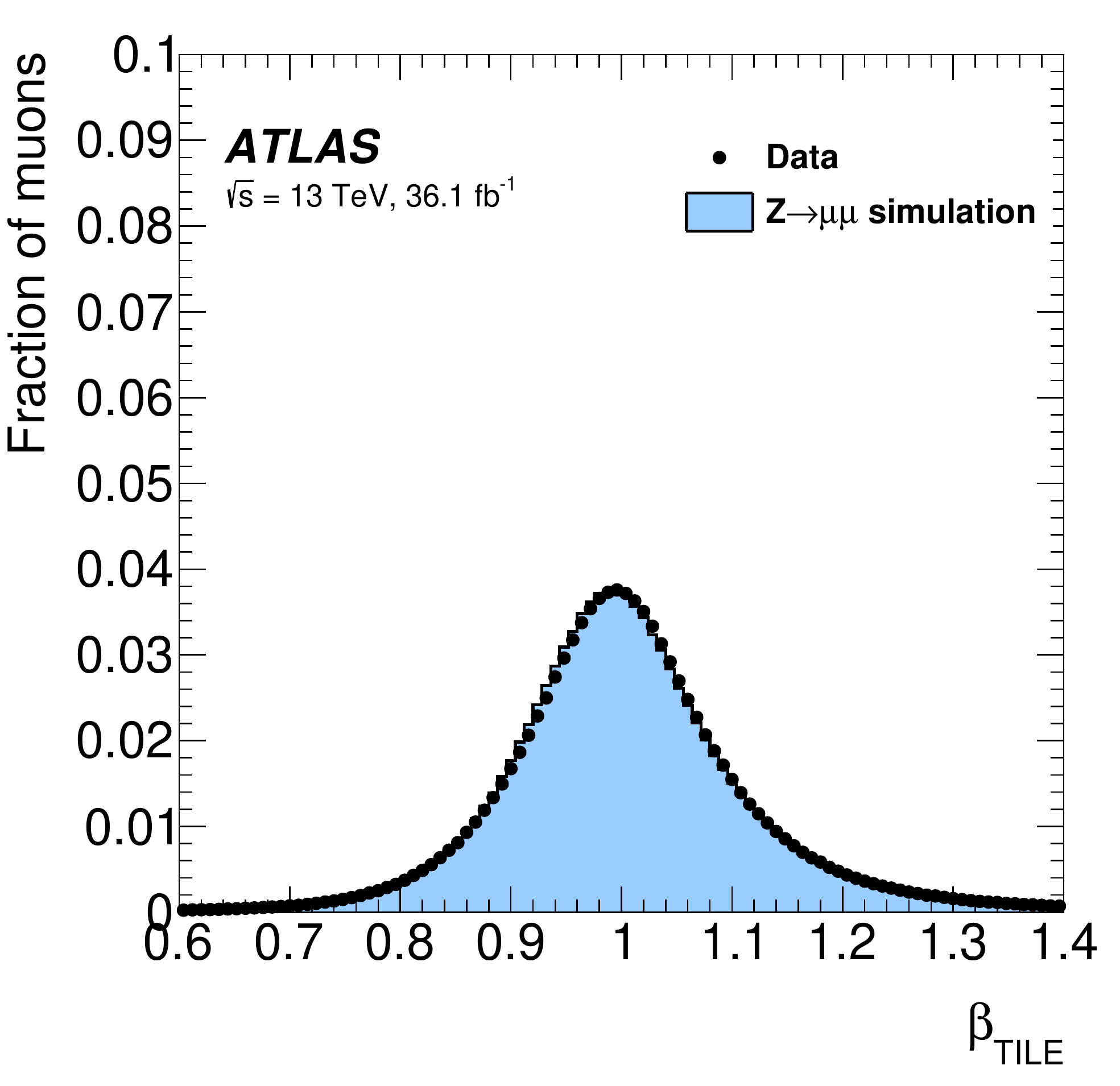}
  \includegraphics[width=.49\textwidth]{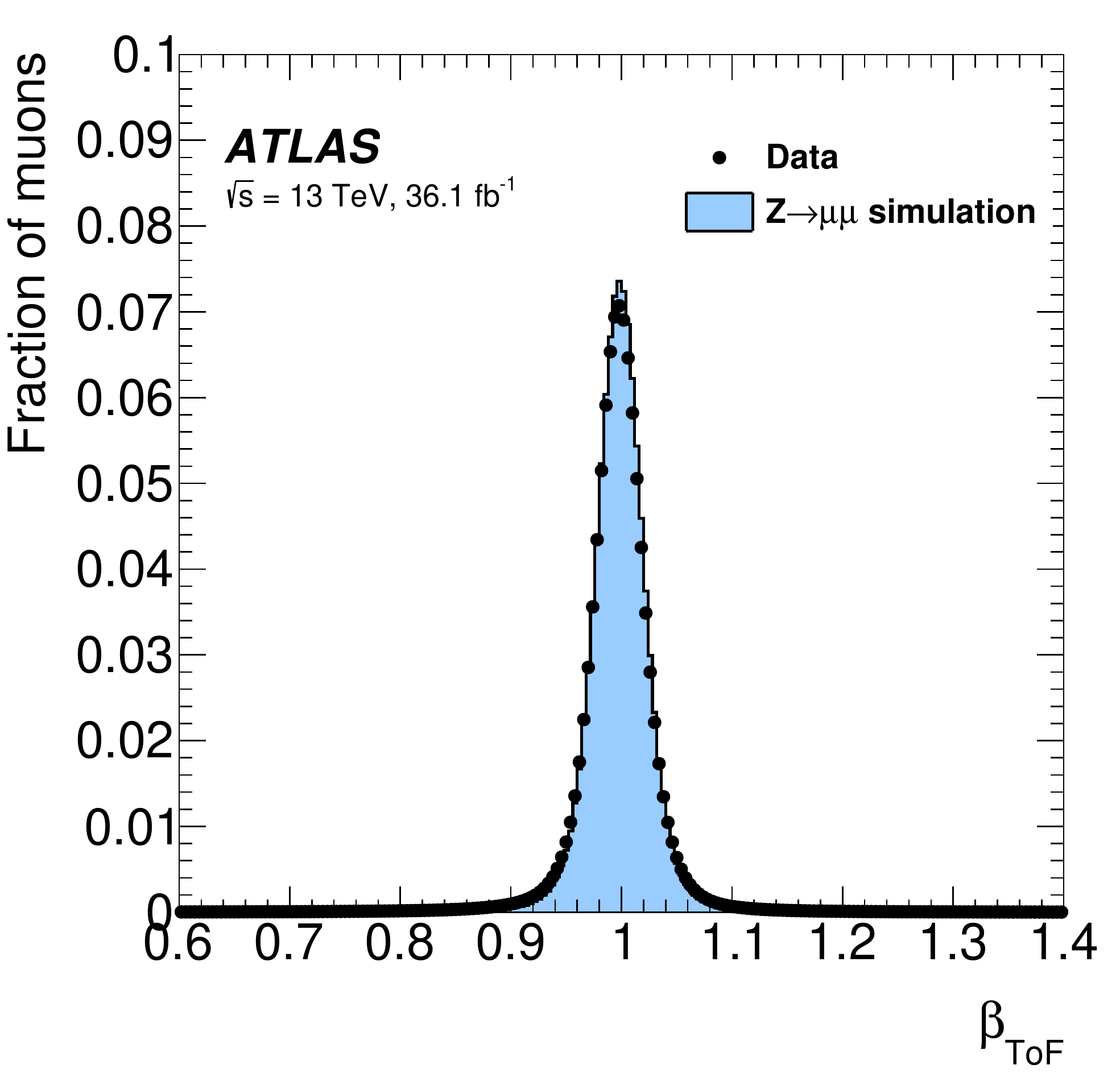}
  \caption{The $\beta$-distributions of muons for a \zmumu selection in data and simulation, with $\beta$ measured solely in the tile calorimeter (\betatile, left), and as a combined measurement from RPCs, MDTs and tile calorimeter (\betatof, right).}
  \label{fig:finalBeta}
\end{figure}

\section{Data and simulated events}
\label{sec:samples}

The analysis presented in this paper is based on a data sample of \lumi of $pp$ collisions at $\sqrt{s} = 13~\TeV$ collected in 2015 and 2016, with a bunch spacing of 25~ns. Reconstructed \zmumu events in data and simulation are used for timing resolution studies. Simulated signal events are used to study the expected signal behavior.

Pair production of gluinos (squarks), with masses between 400~\GeV and 3000~\GeV (600~\GeV and 1400~\GeV), was simulated in \pythia~\cite{Sjostrand:2006za} (version 6.427) with the AUET2B~\cite{ATL-PHYS-PUB-2011-014} tuned set of underlying event and hadronization parameters (tune) and the CTEQ6L1~\cite{Pumplin:2002vw} PDF set, incorporating specialized hadronization routines~\cite{rasmusthesis,Kraan:2004tz} to produce final states containing \rhads. The other SUSY particle masses, except that of the lightest neutralino, were set to very high values to ensure negligible effects on gluino or squark pair-production. The fraction of gluino-balls, i.e.\ bound states of a gluino and gluon, was conservatively set to 10\%~\cite{rasmusthesis, Fairbairn:2006gg}, in order to account for the possibility of other (neutral and hence invisible) final states than those searched for. While the search is optimized for \rhads long-lived enough to reach at least the hadronic calorimeter, samples with gluino lifetimes of 10, 30 and 50~ns, where a significant fraction of the LLPs will decay before the calorimeter, are also investigated.

\pythia relies on a parton shower to add additional high-\pt partons to the event. To achieve a more accurate description of QCD radiative effects, the \pythia events were reweighted to match the transverse-momentum distribution of the gluino--gluino or squark--squark system to the distribution obtained in dedicated leading-order \madgraph (version 2.2.3)~\cite{Alwall:2014hca} simulations with one additional parton in the matrix-element calculation.

Pair production of staus in a GMSB scenario ($m_\text{Messenger}=500~\TeV$, $C_\text{grav}=100000$, $\tan\beta=10$, $\text{sign}(\mu)=1$) was simulated with two additional partons at leading order, using \madgraph (version 2.3.3) in combination with \pythiapp~\cite{Sjostrand:2007gs} (version 8.212) and EvtGen~\cite{Lange:2001uf} (version 1.2.0) with the A14~\cite{ATL-PHYS-PUB-2014-021} tune and NNPDF23LO~\cite{Ball:2012cx} PDF set. The pair-produced lightest stau mass eigenstate is a mixture of the left-handed and right-handed superpartners of the $\tau$-lepton, although it is the partner of the right-handed lepton at the 99\% level in the model considered here.

Pair production of charginos in an mAMSB scenario ($m_0=5~\TeV$, $\tan\beta=5$) was simulated with two additional partons at leading order, using \madgraph (version 2.3.3) in combination with \pythiapp (version 8.212) and EvtGen (version 1.2.0) with the A14 tune and NNPDF23LO PDF set.

Samples of \zmumu events, which are used for general testing, calorimeter and MS timing calibrations as well as the evaluation of systematic uncertainties, were simulated using \powheg~\cite{Alioli:2010xd} (r2856) in combination with \pythiapp (version 8.186) and EvtGen (version 1.2.0) with the AZNLO~\cite{Aad:2014xaa} tune and CTEQ6L1 PDF set.

All events were passed through a full detector simulation~\cite{Aad:1267853} based on the \geant framework~\cite{Agostinelli:2002hh}. For the \rhad simulations, hadronic interactions with matter were handled by dedicated \geant~\cite{Agostinelli:2002hh} routines based on different scattering models: the model used to describe gluino (squark) \rhad interactions is referred to as the generic (Regge) model~\cite{Mackeprang:2009ad}. The \rhads interact only moderately with the detector material, as most of the \rhad momentum is carried by the heavy gluino or squark, which has a small interaction cross section. Typically, the energy deposit in the calorimeters is less than 10~\GeV.

All simulated events included a modeling of contributions from pileup by overlaying minimum-bias $pp$ interactions from the same (in-time pileup) and nearby (out-of-time pileup) bunch crossings simulated in \pythiapp (version 8.186) and EvtGen (version 1.2.0) with the A2~\cite{ATL-PHYS-PUB-2012-003} tune and MSTW2008LO~\cite{Martin:2009iq} PDF set. The simulated events were reconstructed using the same software used for collision data, and were reweighted so that the distribution of the number of collisions per bunch crossing matched that of the data.

\section{Event selection}
\label{sec:selection}

Five dedicated signal regions (SRs), imposing requirements on the entire event and on the individual candidate tracks, are defined in this section, addressing differences in topology and expected interactions with the detector for three different benchmark scenarios: staus, charginos and \rhads.

Events are selected by online triggers based on large missing transverse momentum (\ptmiss, with magnitude denoted by \etmiss) or signatures of single (isolated) high-momentum muons. Large \etmiss values are produced mainly when QCD initial-state radiation (ISR) boosts the \rhad system, resulting in an imbalance between ISR and \rhads whose momenta are not fully accounted for in the \etmiss calculation. The adopted triggers impose thresholds from 70~\GeV to 110~\GeV on \etmiss and 20~\GeV to 26~\GeV on single muons, depending on the data-taking period.

The offline event selection requires all relevant detector components to be fully operational, a primary vertex (PV) built from at least two well-reconstructed charged-particle tracks, each with $\pt > 400~\MeV$, and at least one candidate track that meets the criteria specified below.

A `common track selection' is implemented for candidates in all SRs and is described below. ID tracks (denoted by the superscript `trk') are required to have a minimum $p_\mathrm{T}^\mathrm{trk}$ of 50~\GeV and a momentum measurement $p^\mathrm{trk} < 6.5~\TeV$. The candidate track is required to be matched to the PV using `loose' requirements on the transverse ($d_0$) and longitudinal ($z_0$) impact parameters ($|d_0| < 2$~mm, $|z_0\sin\theta| < 3$~mm).\footnote{The transverse impact parameter is defined as the distance of closest approach between a track and the beam-line in the transverse plane. The longitudinal impact parameter corresponds to the $z$-coordinate distance between the primary vertex and the point along the track at which the transverse impact parameter is defined.} To ensure good track reconstruction, the candidate track must have at least seven silicon clusters\footnote{The charge released by a moving charged particle is rarely contained within just one pixel; neighboring pixels/strips registering hits are joined together using a connected component analysis~\cite{Rosenfeld:1966:SOD:321356.321357, Aad:2014yva} to form clusters.} ($N_\mathrm{silicon}^\mathrm{clusters} > 6$), no shared or split clusters in the pixel detector ($N_\mathrm{pix}^\mathrm{shared}+N_\mathrm{pix}^\mathrm{split} = 0$)~\cite{Aaboud:2017all}, and at least three clusters\footnote{This count includes the number of non-functional/dead modules traversed by the particle/track.} in the SCT ($N_\mathrm{SCT}^\mathrm{clusters+dead} > 2$). Mainly to ensure a reliable timing measurement in the calorimeters, it is required that the sum of the track-\pt in a cone of $\Delta R = 0.2$ around the candidate track is below 5~\GeV. Jets reconstructed in the calorimeter are used to veto electrons and SM hadrons. Jets are reconstructed with the anti-$k_t$ clustering algorithm~\cite{jetalgo} with radius parameter $R = 0.4$ and using as inputs clusters of energy deposits in the calorimeter, calibrated such that the average response of an electron is unity. The jets are then calibrated using the method described in Ref.~\cite{Aaboud:2017jcu}. An electron veto is imposed by rejecting any candidate track for which the nearest jet with $\pt>20~\GeV$ and within a $\Delta R = 0.05$ cone around the track has at least 95\% of its energy deposited in the electromagnetic calorimeter. A veto against SM hadrons is imposed by rejecting any candidate track for which any associated jet within a $\Delta R = 0.05$ cone of the track has an energy larger than the track momentum. Candidate tracks are required to have a cluster in the innermost pixel-detector layer (if expected) or a cluster in the second layer (if the innermost is not expected, but the second layer is). A `cosmics veto' rejects candidate tracks that have an opposite-sign track/muon on the other side of the detector, satisfying $\Delta R_\text{cosmics} < 0.04$.\footnote{$\Delta R_\text{cosmics} = \sqrt{\left(\eta_1 + \eta_2\right)^2 + \left(\Delta\phi\ - \pi \right)^2}$, with $\eta_{1/2}$ being the pseudorapidity of and $\Delta\phi$ the azimuthal angle between the particles in question.} A `$Z$ veto' rejects each candidate track that together with the highest-\pt muon in the event forms an invariant mass within 10~\GeV of the $Z$-boson mass.

Three different selection criteria for each track in an event, hereafter referred to as \selidcalo, \selloose and \seltight, are defined. The \selidcalo selection, which does not use MS information, is designed for the \rhad searches. The \selloose and \seltight criteria are applied to ID--MS combined tracks in the search using the full detector information, with \seltight providing stricter requirements.

An \selidcalo candidate selection starts from an ID track that fulfills the requirements of the above-mentioned `common track selection', and which has at least one timing measurement in the tile calorimeter. In addition, the candidate tracks need to have $|\eta|<1.65$, to ensure a minimal sensitivity in the tile calorimeter. To obtain a reliable \dedx measurement in the pixel detector, the tracks are required to have at least two clusters used in the respective estimation ($N_{\text{good~}\dedx}^\mathrm{clusters} > 1$), a \dedx value between 0 and 20~\MeVgcm, as well as an estimate of $\beta\gamma$ ($0.2 < \bgdedx < 10$). To ensure a reliable \tof measurement in the calorimeter, individual \betatilehit estimates and their individual uncertainties from calorimeter cells associated with the track are combined using a $\chi^2$, and the resulting weighted average \betatile is required to be between 0.2 and 2, with an uncertainty $\sigma_{\betatile} < 0.06$ and a $\chi^2$ probability $P(\chi^2, \mathrm{NDF}) > 0.01$.

A \selloose candidate selection starts from a combined ID--MS track with at least one timing measurement in the tile calorimeter or MS, and fulfilling the requirements of the above-mentioned `common track selection' for its ID track. In addition, the candidate tracks need to have a large combined transverse momentum ($p_\mathrm{T}^\mathrm{cand} > 70~\GeV$), a combined momentum $p^\mathrm{cand} < 6.5~\TeV$, $|\eta|<2$, and hits in at least two MS stations. All \tof-based $\beta$ measurements must be consistent, i.e., for candidates with $\beta$ measurements in more than one \tof system ($N_\text{systems} > 1$), the weighted means of all systems have to be consistent to within 5$\sigma$; and for candidates with only one system ($N_\text{systems} = 1$) the respective internal measurements have to be consistent to within 5$\sigma$. In cases where a \dedx measurement from the pixel detector exists, the $\beta$ estimates based on \dedx and \tof have to be consistent to within 5$\sigma$. The uncertainty in the final $\beta$, whether from a single system or a weighted average of several systems, has to be below 0.025 ($\sigma_{\betatof} < 0.025$) for the candidate to be accepted. At least one system has to yield a $\beta$ measurement and a final value for \betatof between 0.2 and 2 is required.

A \seltight candidate selection is identical to the \selloose selection, except for a tighter pseudorapidity requirement ($|\eta|<1.65$), an additional requirement on the pixel \dedx measurement ($1 < \dedx < 20~\MeVgcm$), and requiring at least two systems to yield a $\beta$ measurement.

To target the three different benchmark scenarios this analysis uses five distinct selections, as shown in Table~\ref{tab:srdefinitions}. Signal regions are defined by imposing requirements on estimated masses, in addition to the criteria in Table~\ref{tab:srdefinitions}. The signal regions that are combined are designed to be orthogonal to allow for a combination in the statistical interpretation of the results.

The search for stable \rhads is performed in both an \msagno SR (\srrhma) and a full-detector SR (\srrhfd) approach. The former is much less dependent on the hadronic-interaction model for \rhads. It is based on events solely selected through \etmiss triggers and candidates stemming from ID tracks fulfilling the \selidcalo selection plus the final selection requirements stated in Table~\ref{tab:srdefinitions}. The full-detector search, on the other hand, takes full advantage of the MS, both in terms of triggering events and in using additional \tof measurements in the MDTs and RPCs.

Candidates originate from combined ID--MS tracks fulfilling the \selloose selection and passing the final requirements states in Table~\ref{tab:srdefinitions}. If no \selloose candidate is found in the event, then it is checked whether there is any candidate satisfying the \selidcalo requirements with the additional selections of Table~\ref{tab:srdefinitions}. The two selections are therefore mutually exclusive. In the rare case of events with more than one candidate, the candidate with the highest \pt is chosen to improve background suppression.

\begin{table}[tbp]
  \centering
  \caption{Summary of the five sets of SRs. The trigger as well as the track candidate selection and the number of candidates per event required for the respective SR are given. Also the final selection requirements together with the mass window (one- or two-dimensional) for the final counting are stated. SRs in one block (delimited by horizontal lines) are combined for the statistical interpretation of the results. For \srrhfd, the \selidcalo is used as a fallback only if no \selloose candidates are found in the event; hence the two SRs are mutually exclusive.}
  \label{tab:srdefinitions}
  \setlength\tabcolsep{1pt}
  \begin{tabular*}{\textwidth}{@{\extracolsep{\fill}}lcccccccc}
    \hline
    \hline
    Signal region              &  Trigger      & Candidate  & Candidates &  \multicolumn{5}{c}{Final requirements}                                            \\
                               &               & selection  & per event  & $|\eta|$    &      $p$ [\GeV] &      \betatof &      \bgdedx & Mass         \\
    \hline
    \srrhma                    & \etmiss       & \selidcalo &  $\ge1$    &  $\le1.65$  &    $\ge200$     &  $\le0.75$    &     $\le1.0$ & \tof\&~\dedx \\
    \hline
    \srrhfd                    & \etmiss/$\mu$ & \selloose  &  $\ge1$    &  $\le1.65$  &    $\ge200$     &  $\le0.75$    &     $\le1.3$ & \tof\&~\dedx \\
    \srrhfd                    & \etmiss/$\mu$ & \selidcalo &  $\ge1$    &  $\le1.65$  &    $\ge200$     &  $\le0.75$    &     $\le1.0$ & \tof\&~\dedx \\
    \hline
    \srtwo                     & \etmiss/$\mu$ & \selloose  &  $=2$      &  $\le2.00$  &    $\ge100$     &  $\le0.95$    &            - & \tof         \\
    \srone                     & \etmiss/$\mu$ & \seltight  &  $=1$      &  $\le1.65$  &    $\ge200$     &  $\le0.80$    &            - & \tof         \\
    \hline
    \hline
  \end{tabular*}
\end{table}

The masses are derived from $m = p / \beta\gamma$ using momentum and \tof measurements to give \mtof, and, where applicable, ionization measurements to give \mdedx. The final selection requirements on the masses are then obtained by fitting the reconstructed mass distribution in signal events for each simulated mass hypothesis with a Gaussian function, and taking the mean minus twice the width as the lower requirement on the respective mass hypothesis. The upper requirement on the mass is left open, and hence is constrained only by kinematics to be < 6.5 TeV. For the \rhad SRs the lower requirements are evaluated in the two-dimensional \mtof--\mdedx-plane, while for SRs \srtwo and \srone only the \mtof distribution is used. This choice is based on the fact that especially for low-mass candidates a significant fraction of candidates have $\dedx < 0.945~\MeVgcm$, at which point there is essentially no separation power between the various mass hypotheses. The final lower requirements on the masses are shown in Figure~\ref{fig:lowerMassCuts}.

The searches for pair-produced stable staus and charginos are performed using two orthogonal SRs. The region \srtwo contains events with exactly two candidates fulfilling the \selloose selection and the corresponding final selection requirements stated in Table~\ref{tab:srdefinitions}, while \srone contains events with exactly one candidate, which in this case must satisfy the \seltight selection and the corresponding final selection requirements stated in Table~\ref{tab:srdefinitions}. For the region \srtwo, which has precedence over \srone in the event categorization, the candidate with the lower \mtof is used to derive limits.

A set of 16 `discovery regions' (DRs) is defined for setting model-independent upper limits on cross sections and stating \pzero-values. These DRs are indicated by larger markers in Figure~\ref{fig:lowerMassCuts}. The resulting DRs are: four for \srrhma, four for \srrhfd (combined \selidcalo $+$ \selloose), as well as four each for exclusive \srtwo and \srone regions.

\begin{figure}[tbp]
  \centering
  \includegraphics[width=.49\textwidth]{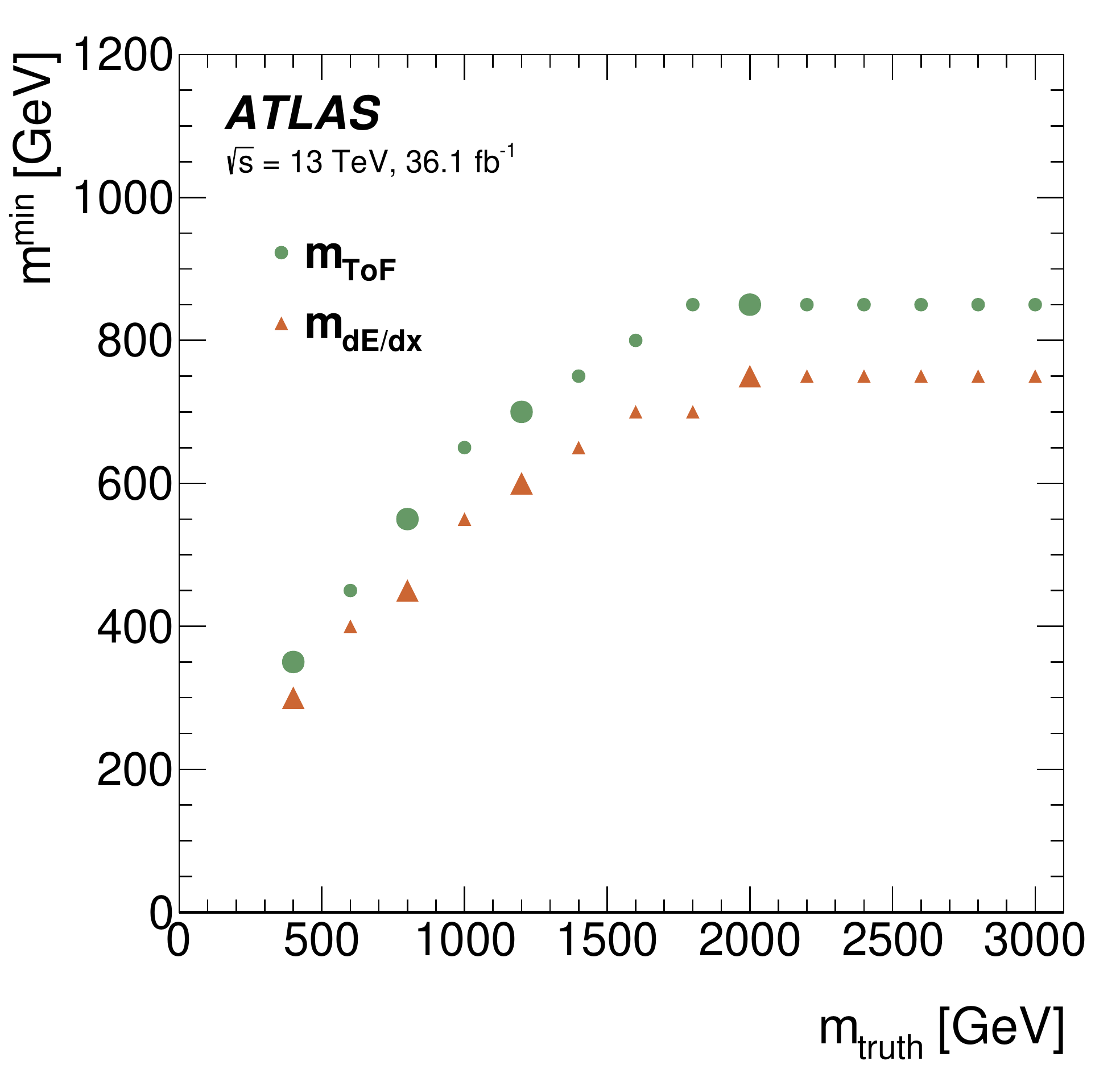}
  \includegraphics[width=.49\textwidth]{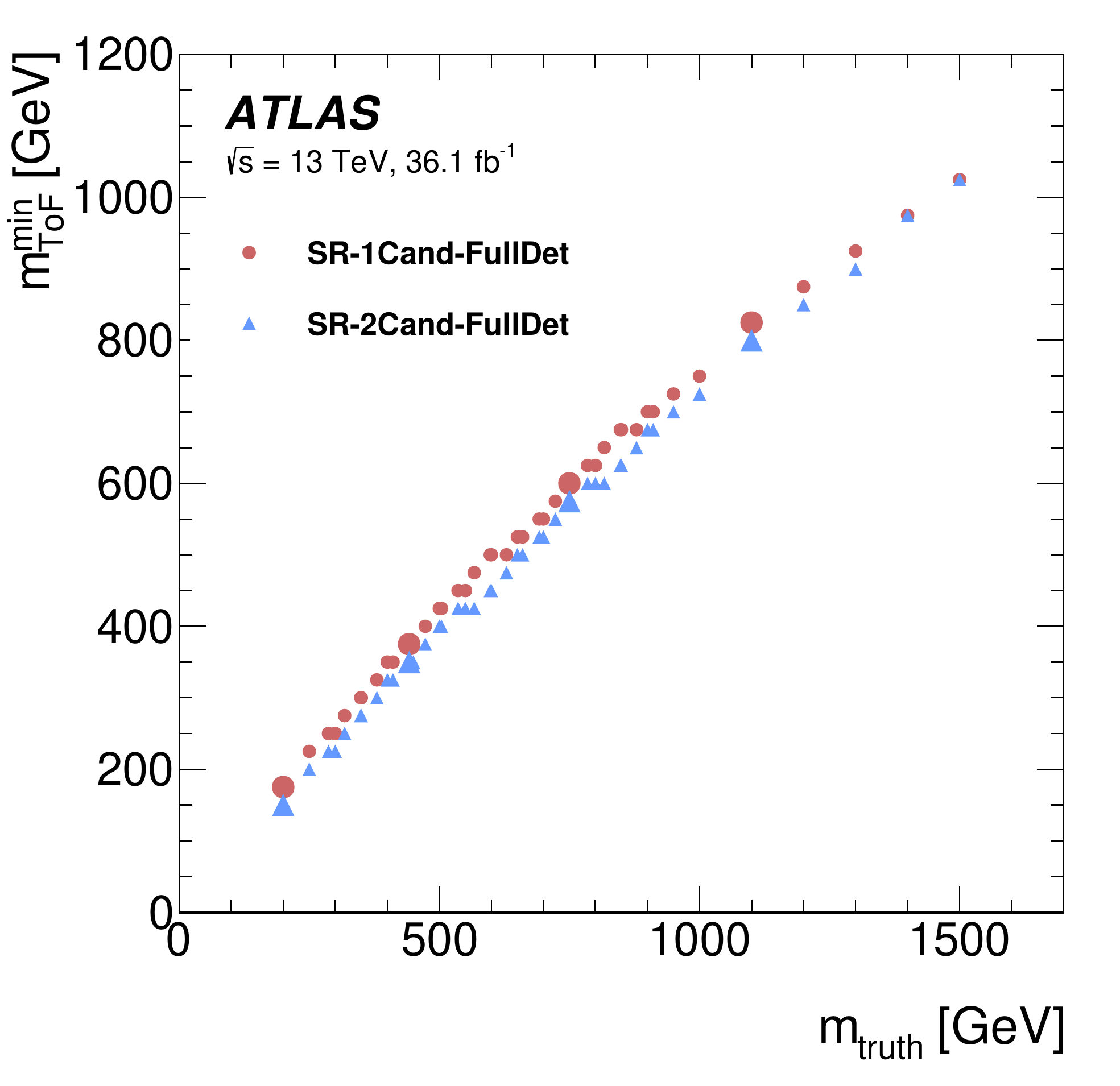}
  \caption{The lower mass requirement ($m_\mathrm{min}$) defining the final counting regions in the \mtof--\mdedx plane used in the \rhad searches (left) and the \mtof-distribution used in the chargino and stau searches (right) for the respective simulated mass of the signal at particle level ($m_\mathrm{truth}$). The larger markers indicate the lower mass requirements for the `discovery regions'.}
  \label{fig:lowerMassCuts}
\end{figure}

\section{Background estimation}
\label{sec:background}

The background is estimated with a fully data-driven method. First, the probability density functions (pdfs) of the key variables are determined from data, using sideband regions where possible. The key variables are momentum, \betatof and \bgdedx for the \rhad SRs, and momentum and \betatof for the chargino and stau SRs. Distributions of expected background in \mtof (and \mdedx) are obtained by randomly sampling the pdfs and using the equation $m=p/{\beta\gamma}$.

For this procedure to be valid, \betatof and \bgdedx must not be correlated with momentum. In principle this is true, as the primary background contribution is from high-momentum muons with mismeasured \betatof and \bgdedx, but there can be an implicit correlation via $\eta$, in particular as the \pt requirement for reconstructing candidates translates into a momentum requirement $p_\text{req} > 200~\GeV$ even for $|\eta| < 2.0$. This means that momentum is correlated with $\eta$ for $|\eta| > 1.75$ due to the lack of low-momentum tracks. As \betatof and \bgdedx are correlated with $\eta$ due to the different resolutions in different detector regions, the result is some correlation of \betatof and \bgdedx with momentum. To remove the effect of these correlations, the pdfs are estimated in five (\srrhma, \srrhfd, \srone) or six (\srtwo) $|\eta|$ bins. A variable binning is used to account for the different regions of the subsystems used. For the sampling of the background, $\eta$ from the candidate is used to get the corresponding pdfs. This is a safe procedure as effects from signal contamination are negligible.

To derive the \srrhma and \srrhfd momentum pdfs, the final requirements on \betatof and \bgdedx are inverted, while for the \betatof and \bgdedx pdfs the final requirement on the momentum is inverted and a minimum momentum of 50~\GeV is required. For the \srtwo, the momentum pdf can be estimated from a sideband in \betatof, but not the \betatof pdf, because in the high-$|\eta|$ region no candidates with $p < 100~\GeV$ are left. \srone instead uses sidebands for the momentum pdfs, since for the tight candidates $|\eta|<1.65$ is required and hence tracks with $p<200~\mathrm{GeV}$ for the full range are available. For the $\beta$ pdf in \srone and \srtwo, a minimum momentum of $70~\mathrm{GeV}$ is required. The final requirements are summarized in Table~\ref{tab:srdefinitions}.

The background yield is normalized to data using low-mass control regions (CRs). The number of estimated events is scaled, so that the number of events in the CR matches the number observed in data. In the \rhad searches, the CR is identical to the SR except that the minimum mass requirement is replaced by an upper limit on \mdedx and \mtof, which both have to be below 300~\GeV. Similarly, in the chargino/stau search, the minimum mass requirement on \mtof is replaced by an upper limit of 150~\GeV and 200~\GeV for \srtwo and \srone, respectively.

The background estimates overlaid with MC signal events at the expected mass limit are shown in Figures~\ref{fig:BKGrhadMAgluino} to \ref{fig:BKGC1C1}. The statistical uncertainty from the pdfs is propagated to the background estimate and shown as gray bands. For all SRs, agreement between data and estimated background in the low-mass regions is found.

\begin{figure}[tbp]
  \centering
  \includegraphics[width=.6\textwidth]{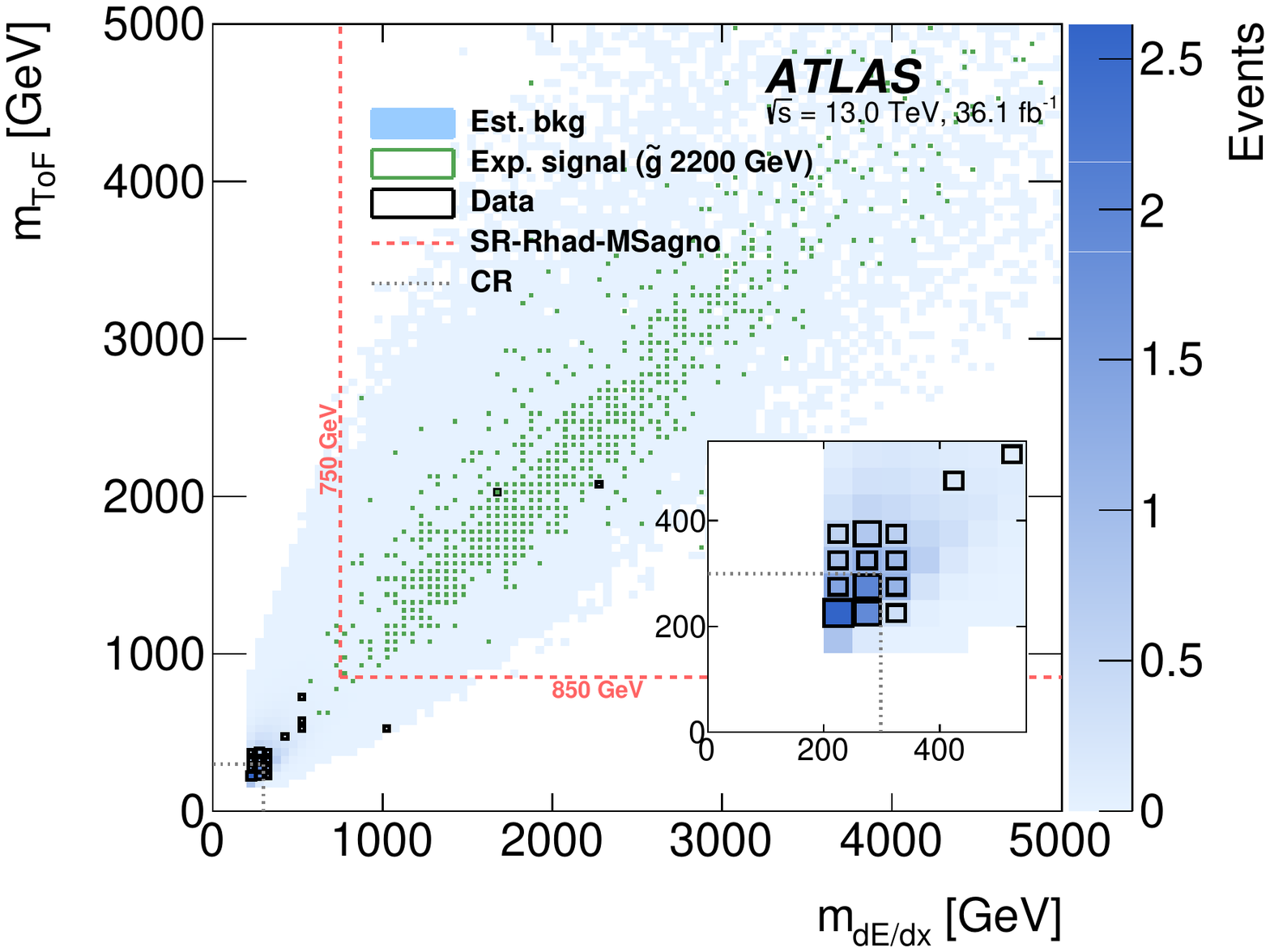}\\
  \includegraphics[width=.48\textwidth]{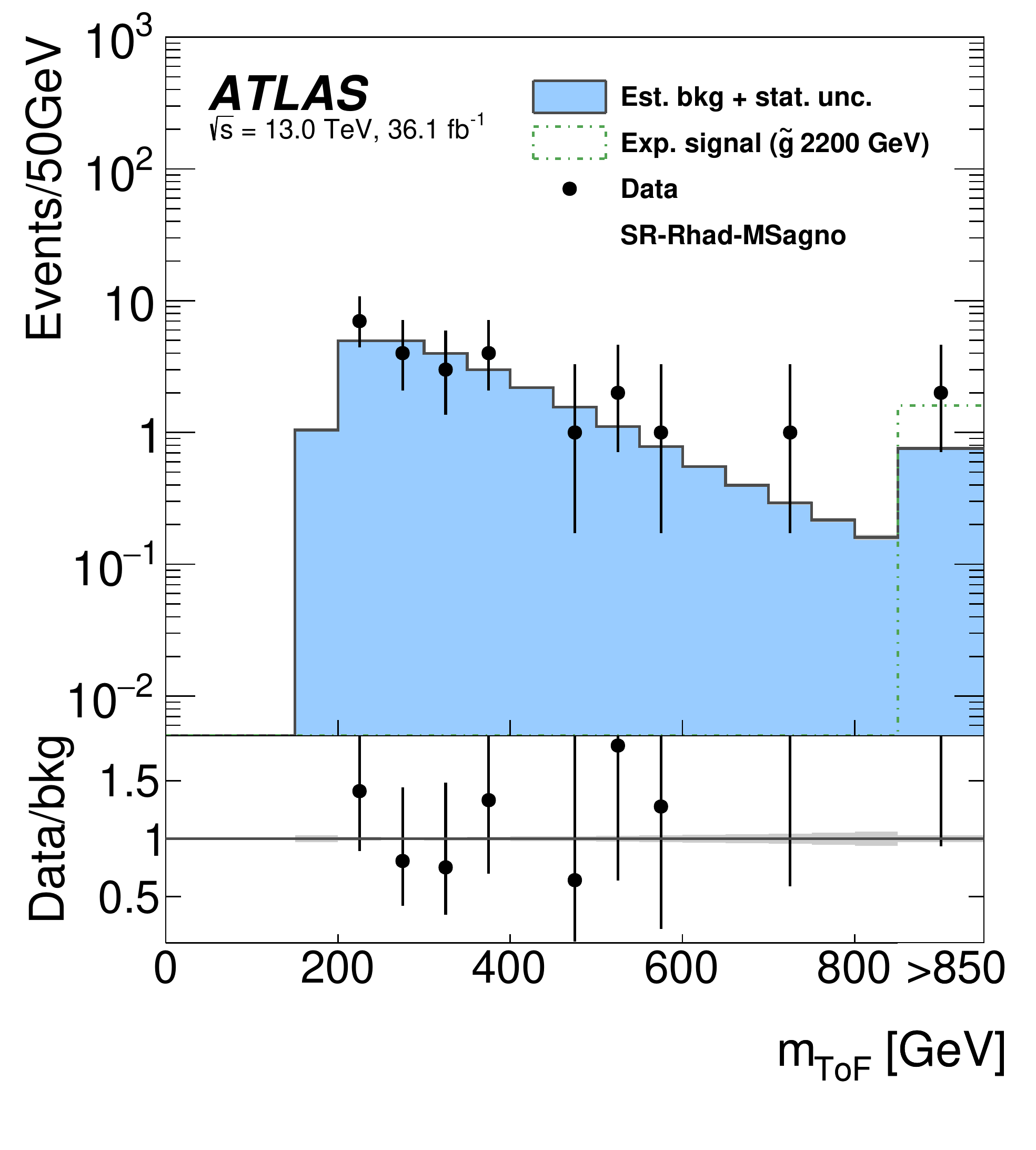}
  \includegraphics[width=.48\textwidth]{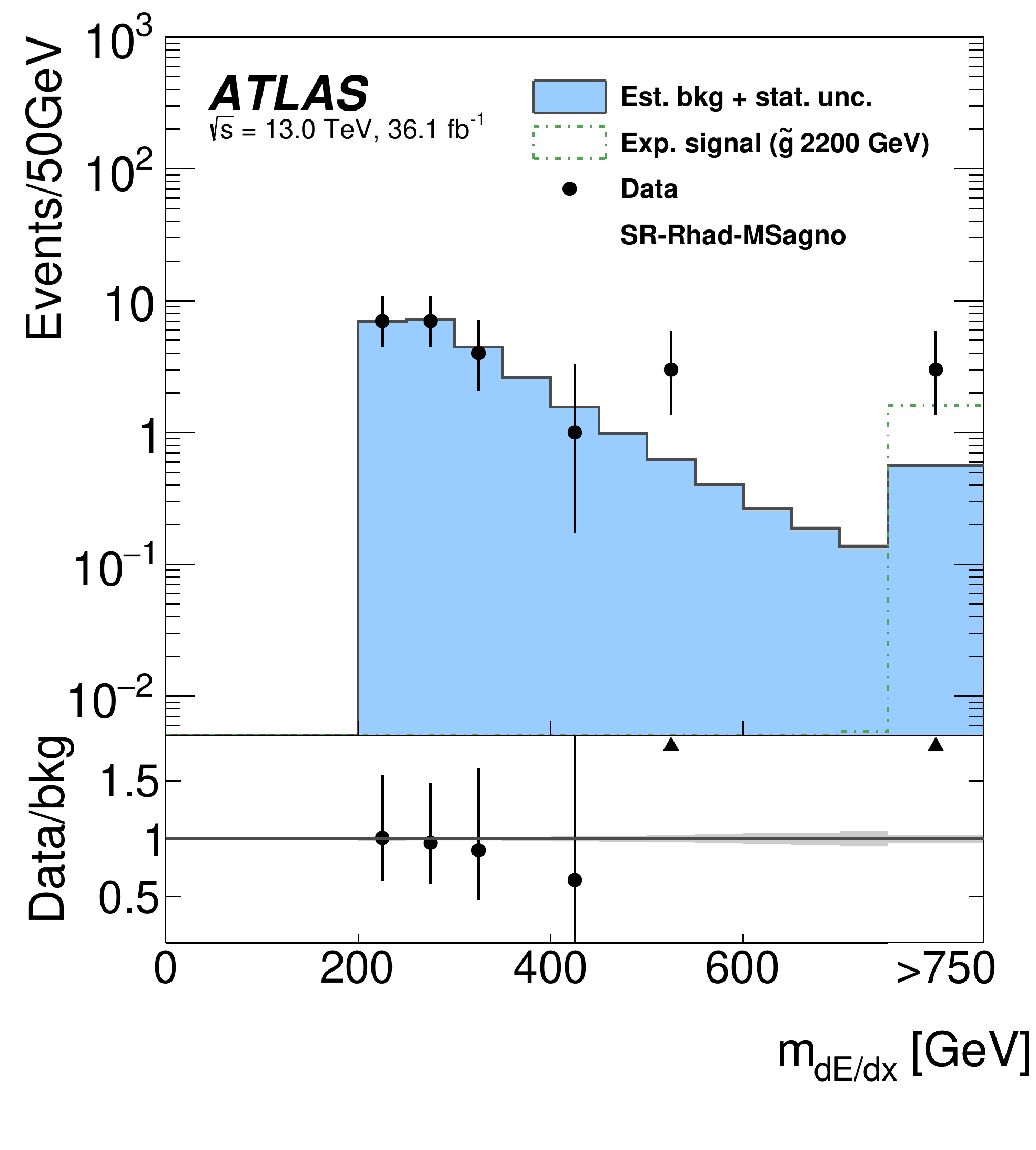}
  \caption{Background estimate for the \msagno analysis targeting gluino \rhads (\srrhma (\selidcalo)) in the \mtof--\mdedx plane (top), the \mtof-projection (bottom-left) and the \mdedx-projection (bottom-right). The last bin of each distribution includes the overflow. The dashed red lines in the upper figure indicate the lower bounds of the signal region for a representative signal choice, while the dotted gray lines illustrate the upper bound for the control region. The signal (2200~\GeV gluino \rhad) is indicated by green markers in the upper and green dash-dotted lines in the lower plots. The lower panels show the ratio of observed data to estimated background. The shaded gray area shows the statistical uncertainty of the background estimate.}
  \label{fig:BKGrhadMAgluino}
\end{figure}

\begin{figure}[tbp]
  \centering
  \includegraphics[width=.6\textwidth]{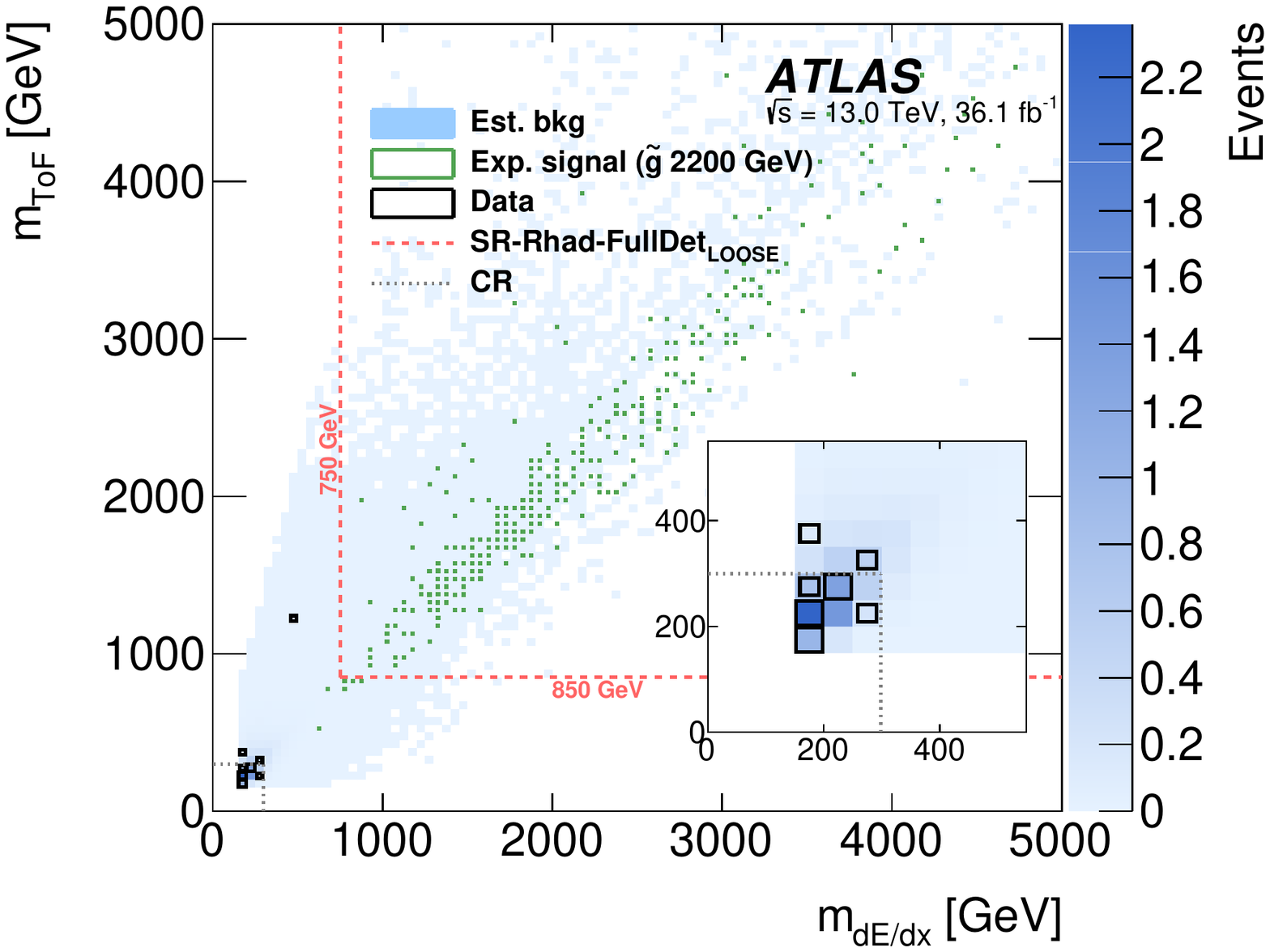}\\
  \includegraphics[width=.48\textwidth]{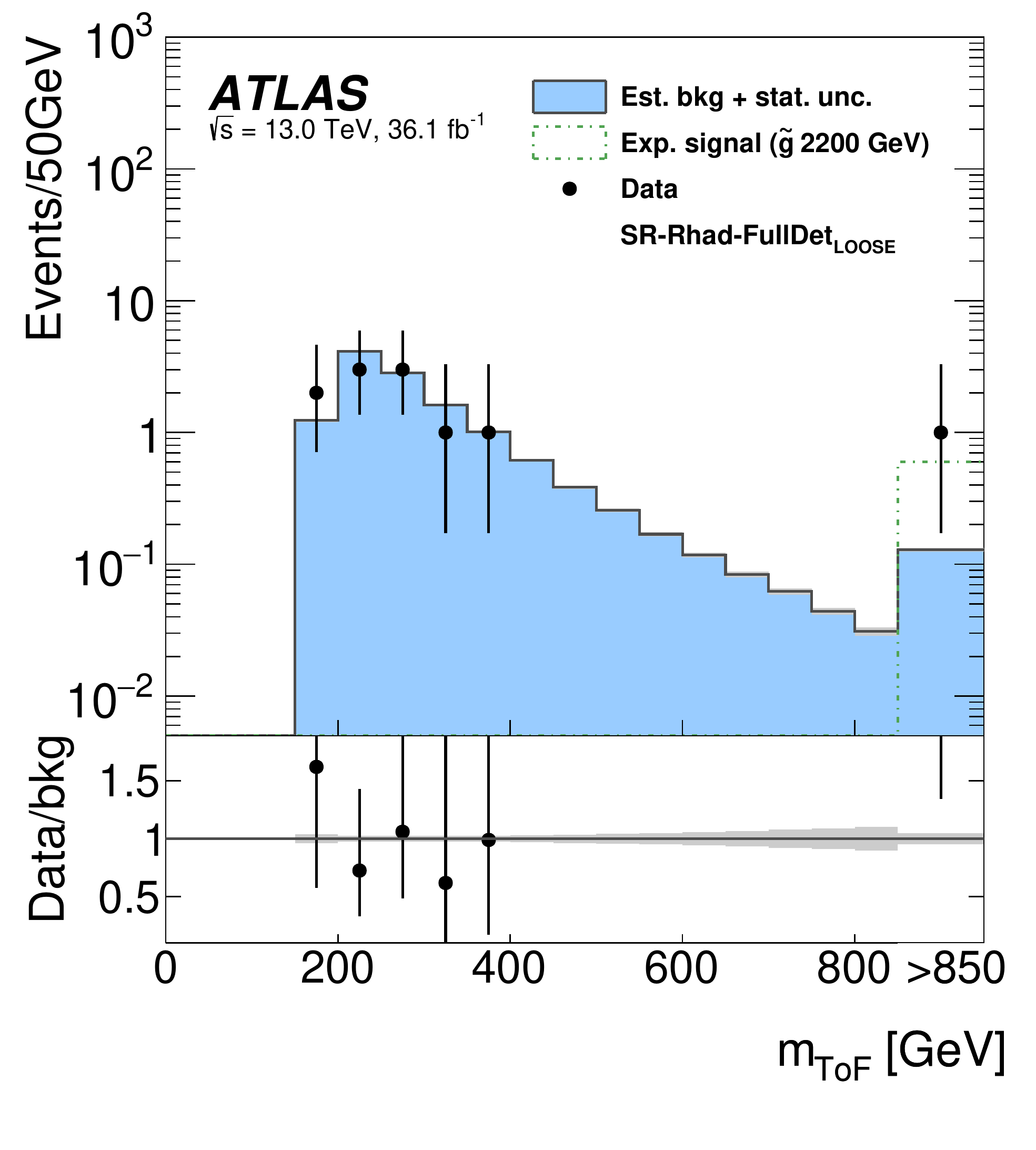}
  \includegraphics[width=.48\textwidth]{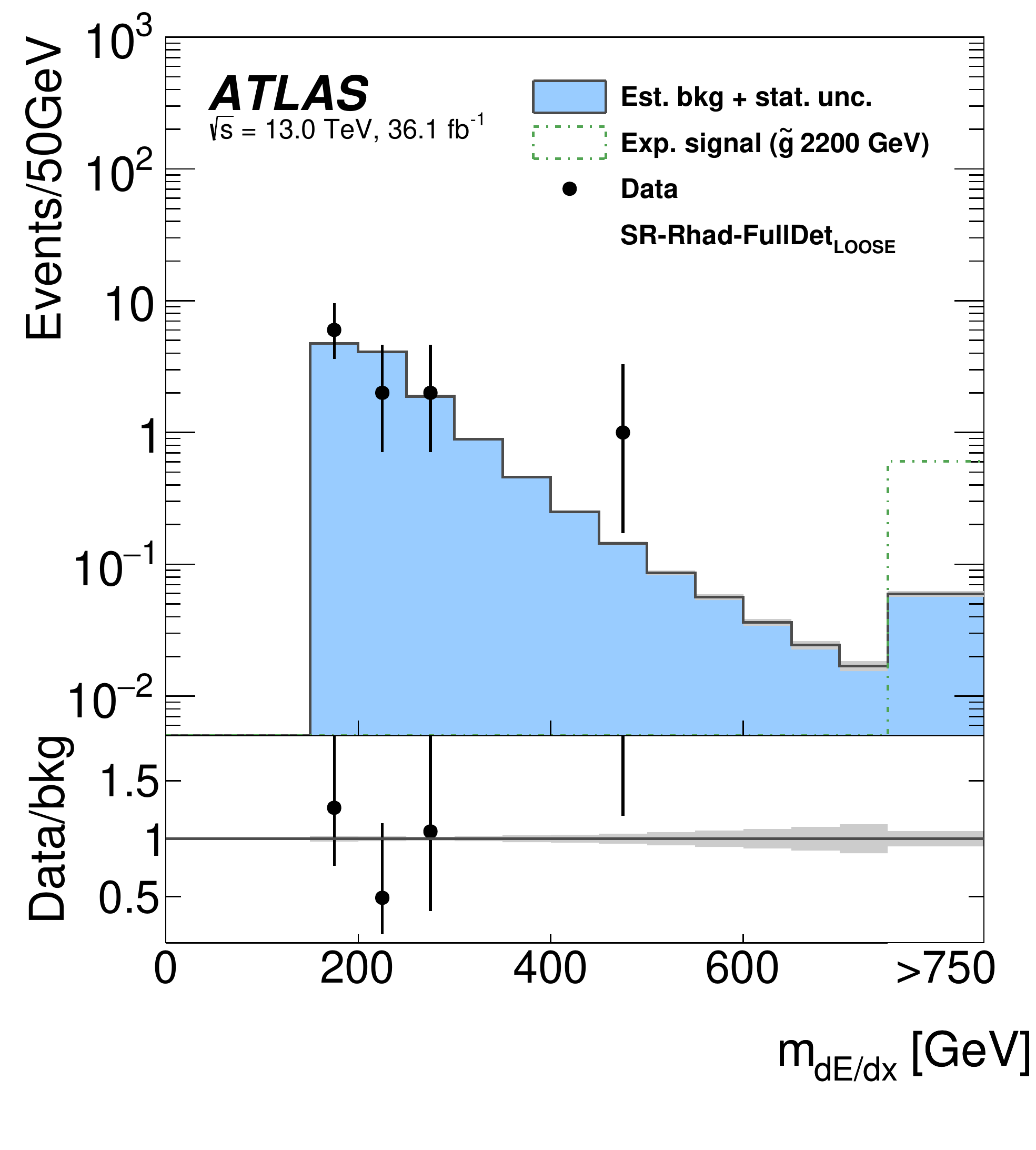}
  \caption{Background estimate for the full-detector analysis (\selloose part) targeting gluino \rhads (\srrhfd) in the \mtof--\mdedx plane (top), the \mtof-projection (bottom-left) and the \mdedx-projection (bottom-right). The last bin of each distribution includes the overflow. The dashed red lines in the upper figure indicate the lower bounds of the signal region for a representative signal choice, while the dotted gray lines illustrate the upper bound for the control region. The signal (2200~\GeV gluino \rhad) is indicated by green markers in the upper and green dash-dotted lines in the lower plots. The lower panels show the ratio of observed data to estimated background. The shaded gray area shows the statistical uncertainty of the background estimate.}
  \label{fig:BKGrhadFDfullgluino}
\end{figure}

\begin{figure}[tbp]
  \centering
  \includegraphics[width=.6\textwidth]{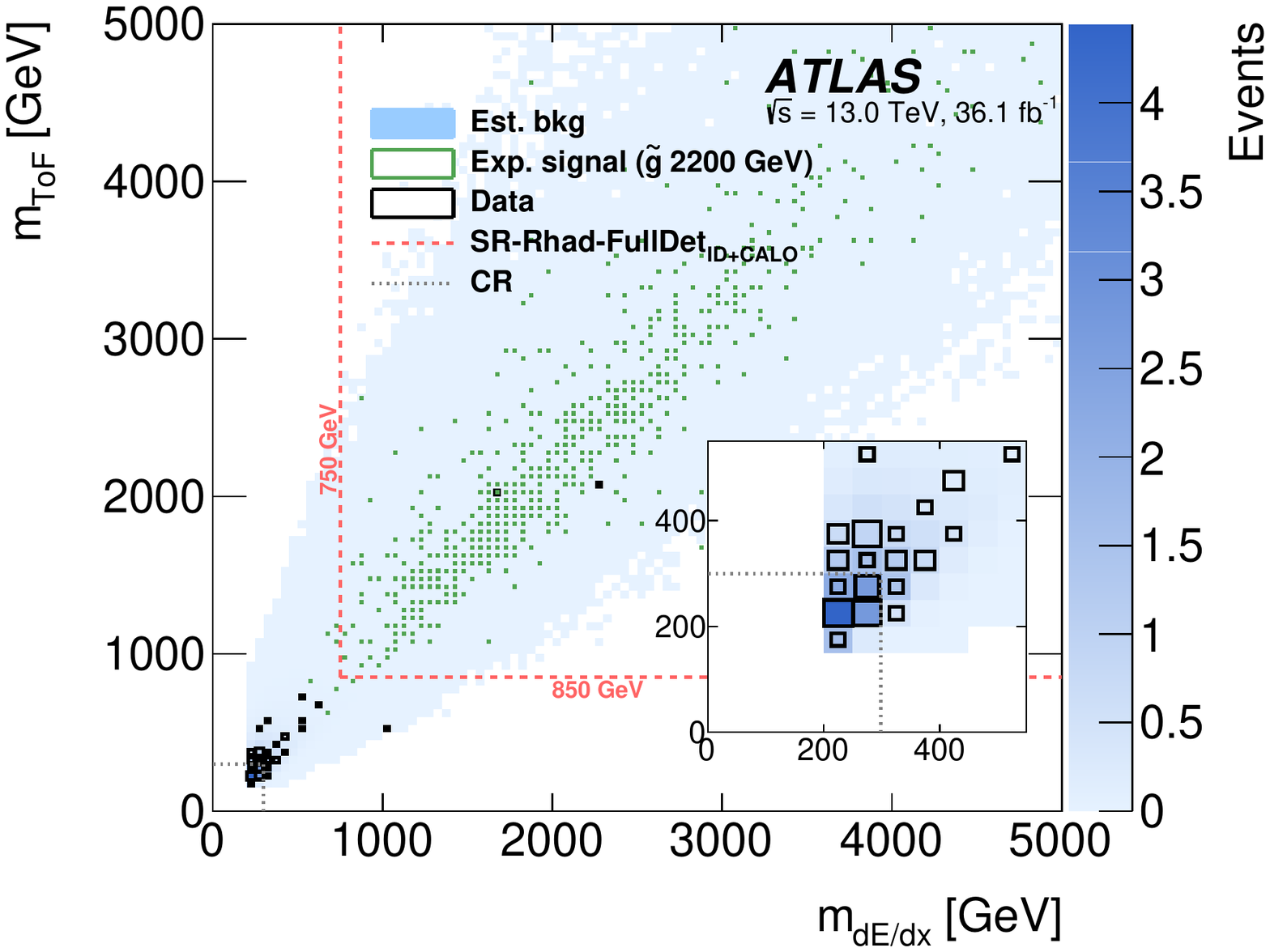}\\
  \includegraphics[width=.48\textwidth]{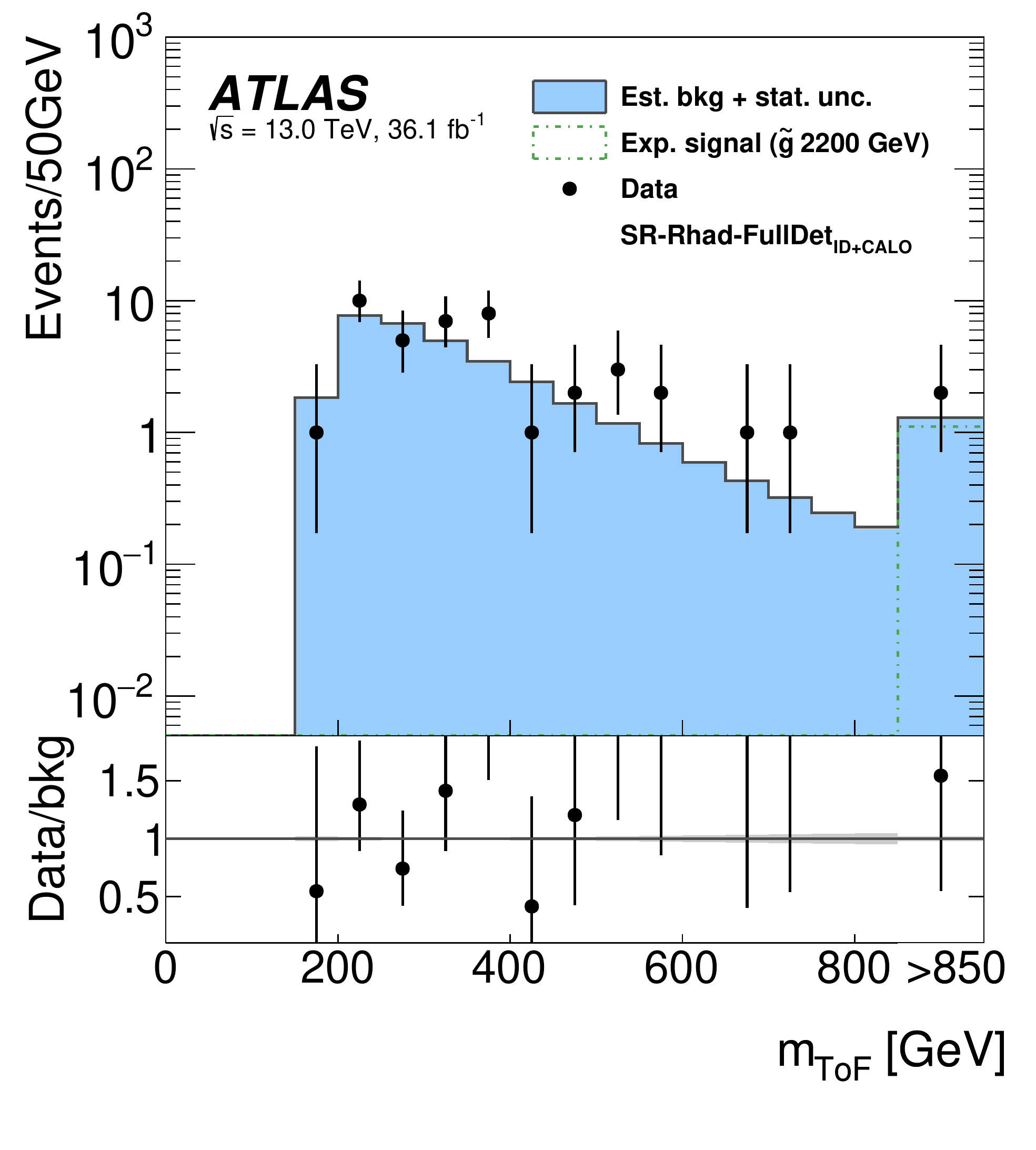}
  \includegraphics[width=.48\textwidth]{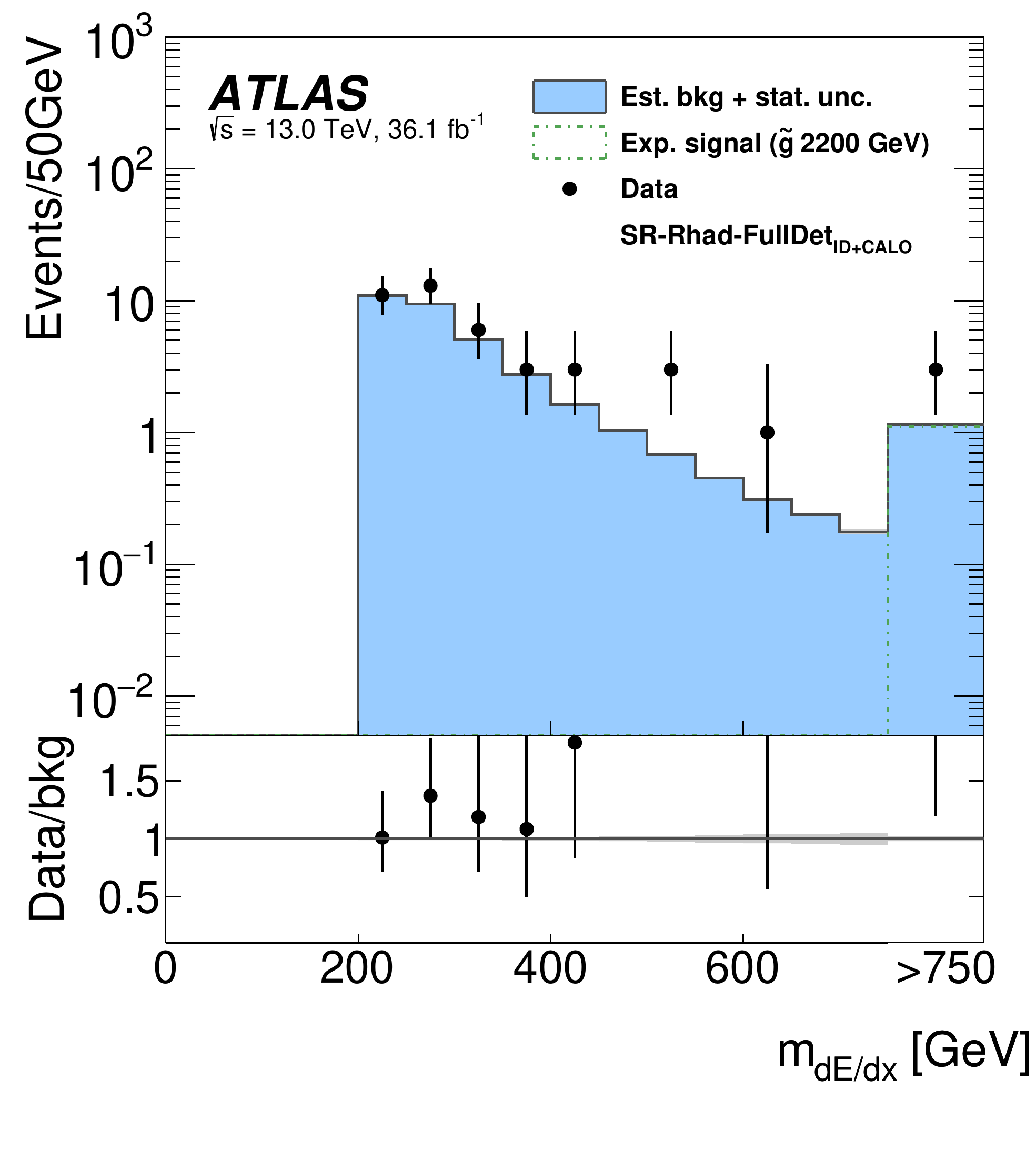}
  \caption{Background estimate for the full-detector analysis (\selidcalo part) targeting gluino \rhads (\srrhfd) in the \mtof--\mdedx plane (top), the \mtof-projection (bottom-left) and the \mdedx-projection (bottom-right). The last bin of each distribution includes the overflow. The dashed red lines in the upper figure indicate the lower bounds of the signal region for a representative signal choice, while the dotted gray lines illustrate the upper bound for the control region. The signal (2200~\GeV gluino \rhad) is indicated by green markers in the upper and green dash-dotted lines in the lower plots. The lower panels show the ratio of observed data to estimated background. The shaded gray area shows the statistical uncertainty of the background estimate.}
  \label{fig:BKGrhadFDIDCalogluino}
\end{figure}

\begin{figure}[tbp]
  \centering
  \includegraphics[width=.48\textwidth]{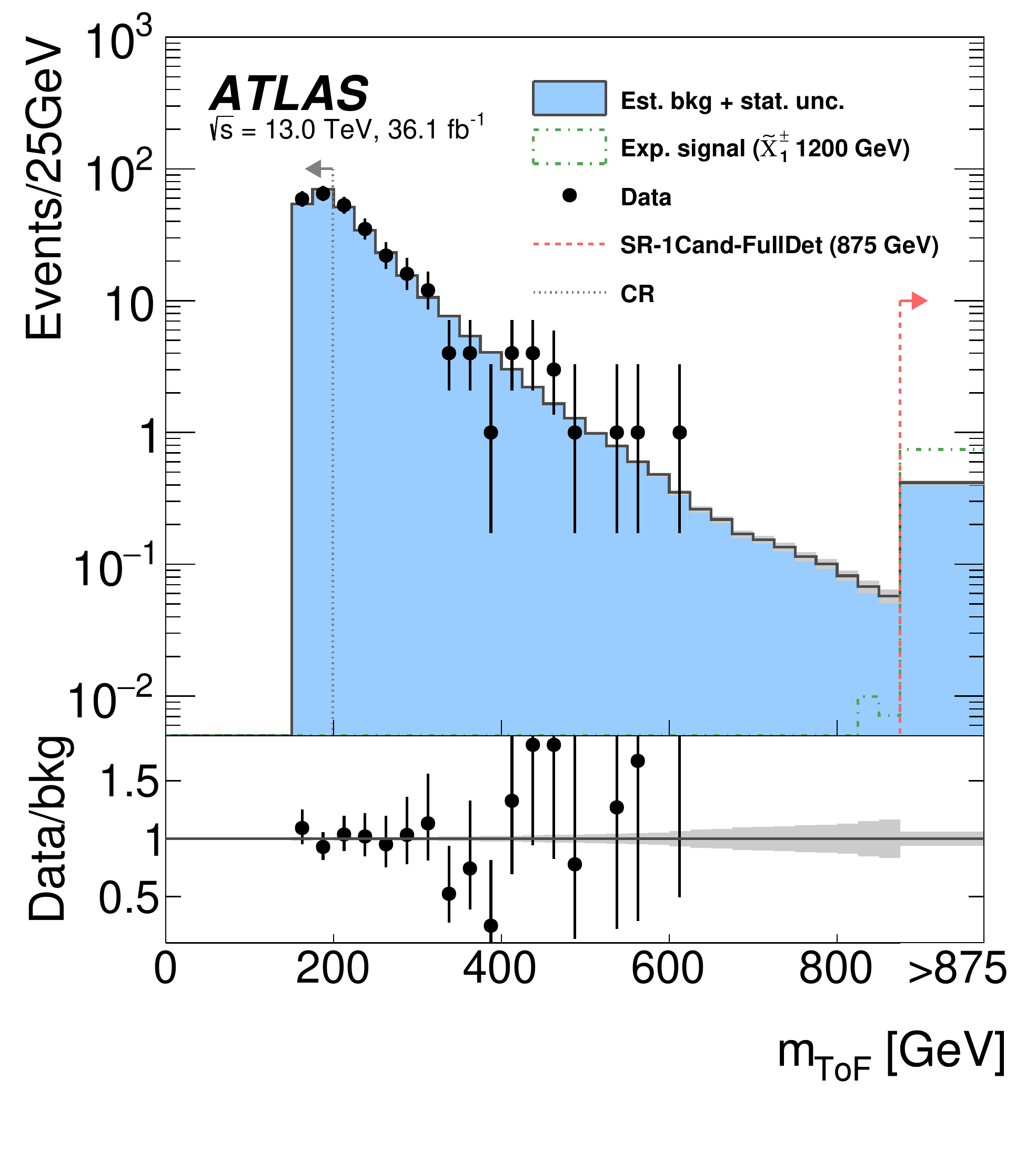}
  \includegraphics[width=.48\textwidth]{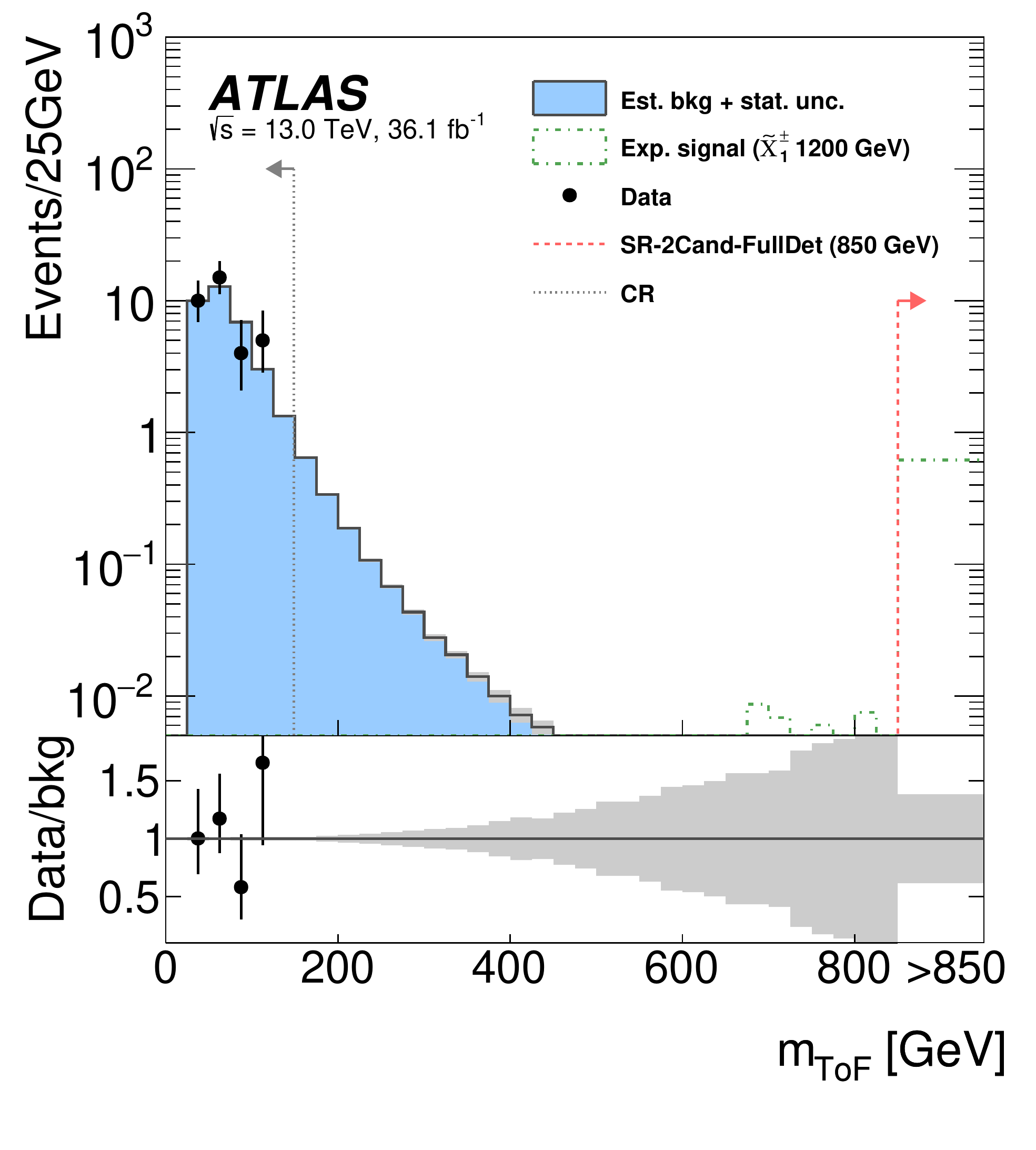}
  \caption{Background estimate for the analysis targeting pair-produced staus and charginos in the one-\seltight-candidate SR (left) and the two-\selloose-candidates SR (right). The last bin of each distribution includes the overflow. The dashed red lines indicate the lower bound of the signal region for a representative signal choice, while the dotted gray lines illustrate the upper bound for the control region. The signal (1200~\GeV chargino) is indicated by green dash-dotted lines. The lower panels show the ratio of observed data to estimated background. The shaded gray area shows the statistical uncertainty of the background estimate.}
  \label{fig:BKGC1C1}
\end{figure}

\section{Systematic uncertainties}
\label{sec:systematics}

The two major uncertainties in the signal yields are from the theoretical cross section and the modeling of ISR, as well as the dedicated full-detector track reconstruction  and the ToF-based $\beta$ measurement in the MS in some cases. All individual contributions are outlined below and summarized in Table~\ref{tab:systsummary}.

\textbf{Theoretical cross sections}

\rhad production cross sections are calculated to next-to-leading order (NLO) in the strong coupling constant, adding the resummation of soft-gluon emission at next-to-leading-logarithm accuracy (NLO+NLL)~\cite{Beenakker:1996ch, Kulesza:2008jb, Kulesza:2009kq, Beenakker:2009ha, Beenakker:2011fu}. Stau and chargino signal cross sections are calculated to NLO in the strong coupling constant (NLO) using \prospino~\cite{Beenakker:1996ed}. The nominal cross section and the uncertainty is taken from an envelope of cross-section predictions using different PDF sets and factorization and renormalization scales, as described in Ref.~\cite{Borschensky:2014cia}. This prescription results in an uncertainty in the cross section of between 14\% (at 600~\GeV) and 57\% (at 3000~\GeV) for gluino \rhads, and between 14\% (at 600~\GeV) and 23\% (at 1400~\GeV) for squark \rhads. For direct pair-production of staus and chargino pair-production the uncertainty is between 6\% (at 290~\GeV) and 10\% (at 910~\GeV) and between 4\% (at 200~\GeV) and 10\% (at 1500~\GeV), respectively.

\textbf{Signal efficiency}

Missing-transverse-momentum triggers used in these searches rely solely on calorimeter-energy deposits to calculate the transverse energy, and are thus largely blind to muons, which can therefore be used for calibration and systematic uncertainties. To evaluate the trigger efficiency, the trigger turn-on curve is obtained by fitting the measured efficiency as a function of \etmiss in \zmumu events, in both data and simulation. These efficiency turn-on curves are then applied to the \etmiss spectrum from simulated events. The total uncertainty is estimated from four contributions: the relative difference between the efficiencies obtained using the fitted threshold curves from \zmumu in data and simulation, the differences in efficiency obtained from independent $\pm1\sigma$ variations in fit parameters relative to the unchanged turn-on-curve fit for both \zmumu data and simulation, and a 10\% variation of the \etmiss to assess the scale uncertainty. The \etmiss trigger is estimated to contribute a total uncertainty of 1.6\% and 2\% to the signal efficiency for \rhads and staus/charginos, respectively.

To account for a possible mismodeling of the single-muon trigger timing in simulation, all simulated events were reweighted, depending on the $\beta$ and $|\eta|$ of the candidate, to match the data and to compute systematic uncertainties. Systematic uncertainties of 1\%, 4\% and 3\% are assigned to \rhads, staus and charginos, respectively, by taking the difference in signal trigger efficiency between unweighted and reweighted events.

To address a possible mismodeling of ISR, and hence \etmiss in the signal events, half of the difference between the selection efficiency for the nominal \pythia events and those reweighted to match \madgraph predictions is taken as an uncertainty in the expected signal yield, and found to be below 15\% in all cases. For staus and charginos, the uncertainties in the amount of ISR (and FSR) are evaluated by varying generator parameters in the simulation. The uncertainty from the choice of renormalizaton/factorization scale is evaluated by changing the default scale by a factor of two in \madgraph. The uncertainty from the choice of CKKW-L merging~\cite{Catani:2001cc, Lonnblad:2011xx} scale is evaluated by taking the maximum deviation from the nominal MC sample when varying it by a factor of two. The uncertainty from the parton-shower generator tuning is evaluated using \pythiapp tune variations. The systematic uncertainty on the ISR/FSR is calculated by adding all three components in quadrature. The uncertainties are evaluated for three mass points of staus and charginos and are estimated to be 4\% and 5\%, respectively. The latter procedure was also implemented for the \madgraph gluino--gluino and squark--squark samples, and the resulting uncertainties are similar to the nominal-vs-reweighted ones.

The uncertainty in the pileup modeling in simulation is found to affect the signal efficiency by between 0.1\% and 5.5\%, typically decreasing as a function of the simulated LLP mass and varying with benchmark model.

An additional systematic uncertainty in the signal efficiency is estimated for the dedicated tracking algorithm to cover all discrepancies between data and simulation, by randomly rejecting 10\% of the reconstructed objects. The effect on the final signal efficiency is found to be between 0.2\% and 14.8\%

Uncertainties in the measurement of \dedx in the pixel detector result in uncertainties in the signal yield ranging from 6\% to 13\% for \rhads and less than 0.5\% for staus and charginos. They account for both the shape difference between the ionization distribution in data and simulation, and the scale shift in data due to radiation damage.

The systematic uncertainty in the calorimeter-based $\beta$ estimation is assessed by scaling the calorimeter-cell-time smearing of simulated events by $\pm$5\% and by varying the cell-time correction introduced to correct for the bias due to the OFA by $\pm$50\%, and is found to be below 2\% in all cases.

The systematic uncertainty in the MS-based $\beta$ estimation is derived by varying the MDT smearing constants by $\pm$10\% to bracket the distribution seen in data. By comparing the signal efficiency with and without the correction for incorrectly modeled timing behavior in the RPCs, the overall uncertainty is found to be between 1.0\% and 21.7\%.

\textbf{Integrated luminosity}

The uncertainty in the combined 2015+2016 integrated luminosity is \lumiUnc. It is derived, following a methodology similar to that detailed in Ref.~\cite{Aaboud:2016hhf}, from a calibration of the luminosity scale using $x$--$y$ beam-separation scans performed in August 2015 and May 2016.

\textbf{Background estimation}

To estimate the systematic uncertainty of the background estimate, three main contributions are considered: the systematic uncertainty of the normalization as well as the influence of the $|\eta|$ binning and the definition of the sidebands. The normalization uncertainty, derived by applying the relative statistical uncertainty in the number of data events in the CR to the estimated number of background events in the SR, is the dominant contribution for most SRs. Only at very high masses do the other contributions become significant and eventually take over. To test the stability of the choice of $|\eta|$ binning, the number of $|\eta|$ bins in the pdfs is varied from nominal five (six) to three (four) and six (nine) for \srrhma, \srrhfd and \srone (\srtwo). For each pdf, the background is estimated and half the maximal difference in the number of background events in the SR is taken as a systematic uncertainty. To check for an influence of the sideband definition, the selection requirements are varied, the pdfs re-evaluated correspondingly, and half the maximal difference of background counts in the SR, using the nominal and varied pdfs, is taken as the systematic uncertainty.

\begin{table}[tbp]
  \centering
  \caption{Summary of systematic uncertainties. Ranges indicate a dependence on the mass hypothesis.}
  \label{tab:systsummary}
  \setlength\tabcolsep{1pt}
  \begin{tabular*}{\textwidth}{@{\extracolsep{\fill}}lrrrr}
    \hline
    \hline
    & \multicolumn{4}{c}{ Relative uncertainty~[\%]} \\
                                               &   \msagno & Full-detector &           &           \\
    Source                                     &    \rhads &        \rhads &     Staus & Charginos \\
    \hline
    Theoretical inclusive cross section        &    14--57 &        14--57 &     6--10 &     4--10 \\
    \hline
    Total uncertainty in signal efficiency     &    17--19 &        18--30 &     7--15 &     9--18 \\
    ~~~Trigger efficiency                      &       1.6 &           1.9 &       4.5 &       3.9 \\
    ~~~~~\etmiss                               &       1.6 &           1.6 &       2.0 &       2.5 \\
    ~~~~~Single-muon                           &           &           1.0 &       4.0 &       3.0 \\
    ~~~Theoretical uncertainty (ISR/FSR)       &        15 &            15 &         4 &         7 \\
    ~~~Pileup                                  &  0.2--3.8 &      0.3--5.5 &  0.1--3.1 &  0.2--4.4 \\
    ~~~Full-detector track reconstruction      &         - &     1.7--14.8 & 0.2--12.8 & 0.8--13.0 \\
    ~~~Track hit requirements                  &         2 &             2 &         2 &         2 \\
    ~~~Pixel $\beta\gamma$ measurement         & 6.0--11.6 &     6.0--13.0 &       0.5 &       0.5 \\
    ~~~ToF $\beta$ measurement                 &  0.5--3.6 &     9.8--21.9 &  1.0--3.6 & 2.0--12.0 \\
    ~~~~~Calorimeter $\beta$ measurement       &  0.1--0.5 &      0.1--1.1 &  0.1--0.5 &  0.1--0.5 \\
    ~~~~~Calorimeter OFA correction            &  0.4--3.6 &      1.2--3.1 &  0.1--0.4 &  0.1--1.3 \\
    ~~~~~MS $\beta$ measurement                &         - &     9.7--21.7 &  1.0--3.5 & 2.0--12.0 \\
    \hline
    Luminosity                                 &       2.1 &           2.1 &       2.1 &      2.1  \\
    \hline
    Uncertainty in background estimate         &    33--34 &        27--53 &     9--31 &    9--34  \\
    \hline
    \hline
  \end{tabular*}
\end{table}

\section{Results}
\label{sec:result}

Mass distributions observed in data together with the background estimate, its statistical uncertainty and a representative expected signal are shown in Figures~\ref{fig:BKGrhadMAgluino} to \ref{fig:BKGC1C1} for the \msagno and full-detector \rhad, stau and chargino searches, respectively.

As can be seen in Tables~\ref{tab:finalnumbers_rhad_ma} to \ref{tab:finalnumbers_chargino}, no significant excess of observed data events above the expected background is found in the examined mass ranges and signal regions. The yields are summarized for a subset of discovery regions (see Section~\ref{sec:selection}) in Figure~\ref{fig:discSummary} and Table~\ref{tab:pzeros}, the latter also showing the \pzero-values.

Upper limits at 95\% confidence level (CL) are placed on the production cross sections for various benchmark models, as shown in Figures~\ref{fig:limitrhadComb} and \ref{fig:limit2Cand1Cand}. These limits are obtained from the expected signal and the estimated background in signal region \srrhma (\srrhfd or \srone/ \srtwo) using a one-bin (two-bin) counting experiment applying the \CLs prescription~\cite{Read:2000ru}. Model-independent upper limits defined as cross section $\times$ acceptance $\times$ efficiency for the above-mentioned discovery regions are shown in Table~\ref{tab:pzeros}. Given the predicted theoretical cross sections, also shown in Figures~\ref{fig:limitrhadComb} and \ref{fig:limit2Cand1Cand}, the cross-section limits are translated into lower limits on masses for the various benchmark models.

The \msagno search yields expected lower limits at 95\% CL on the \rhad masses of \limRHMAgluExp, \limRHMAsboExp and \limRHMAstoExp for the production of long-lived gluino, sbottom and stop \rhads, respectively. The corresponding observed lower limits on the masses are \limRHMAgluObs, \limRHMAsboObs and \limRHMAstoObs. The expected signal yield ($N_\text{exp.}$) and efficiency, estimated background ($N_\text{est.}$) and observed number of events in data ($N_\text{obs.}$) for the full range of simulated masses can be found in Table~\ref{tab:finalnumbers_rhad_ma}. The sensitivity first increases and then decreases with increasing \rhad mass. The same effect is visible in the total efficiency of the \etmiss trigger, and is due to the change in production channel for gluino \rhads from gluon-initiated to quark-initiated with increasing mass.

For metastable gluino \rhads, the \msagno search yields expected lower limits on mass of \limRHMAgluExpTen, \limRHMAgluExpThrity and \limRHMAgluExpFivty for lifetimes of 10~ns, 30~ns and 50~ns, respectively. The corresponding observed lower limits are \limRHMAgluObsTen, \limRHMAgluObsThrity and \limRHMAgluObsFivty. The expected signal yield ($N_\text{exp.}$) and efficiency, estimated background ($N_\text{est.}$) and observed number of events in data ($N_\text{obs.}$) for the full range of simulated masses can be found in Table~\ref{tab:finalnumbers_rhad_ma_meta}. As additional \etmiss can arise from one of the gluino \rhads decaying before the calorimeters, the sensitivity increases for mass regions that have limited trigger efficiency (low and high masses). With decreasing lifetime, and thereby increasing probability for both gluino \rhads to decay before the calorimeters, the signal efficiency drops, as does the sensitivity of the search optimized for \rhads long-lived enough to exit the detector. However, the expected lower limits on mass at lifetimes of 50~ns and 30~ns remain more stringent than those of the search targeting low-lifetime metastable gluino \rhads~\cite{SUSY-2016-31}. Expected and observed lower limits on mass as a function of lifetime for both the \msagno and full-detector searches for gluino \rhads are shown in Figure~\ref{fig:limitsRhadGluinoLifetimeSummary}.

Using the full-detector search, the expected (observed) lower limits on the mass are \limRHFDgluExp (\limRHFDgluObs), \limRHFDsboExp (\limRHFDsboObs) and \limRHFDstoExp (\limRHFDstoObs) for the production of long-lived gluino, sbottom and stop \rhads, respectively. The expected signal yield ($N_\text{exp.}$) and efficiency, estimated background ($N_\text{est.}$) and observed number of events in data ($N_\text{obs.}$) for the full range of simulated masses can be found in Table~\ref{tab:finalnumbers_rhad_fd}. The sensitivity of the \rhad full-detector search in mass regions with noticeable background yields is slightly better than that of the \msagno search, due mainly to the increased signal efficiency when including the single-muon trigger and an improved $\beta$-resolution when using full-detector candidates. However, this specific search is consequently slightly more model-dependent, especially on the modeling of hadronic interactions.

Expected (observed) lower limits on mass for direct production of staus and charginos are set at \limStauExp (\limStauObs) and \limCharExp (\limCharObs), respectively. The expected signal yield ($N_\text{exp.}$) and efficiency, estimated background ($N_\text{est.}$) and observed number of events in data ($N_\text{obs.}$) for the full range of simulated masses can be found in Tables~\ref{tab:finalnumbers_stau} and \ref{tab:finalnumbers_chargino}.

\begin{figure}[tbp]
  \centering
  \includegraphics[width=.95\textwidth]{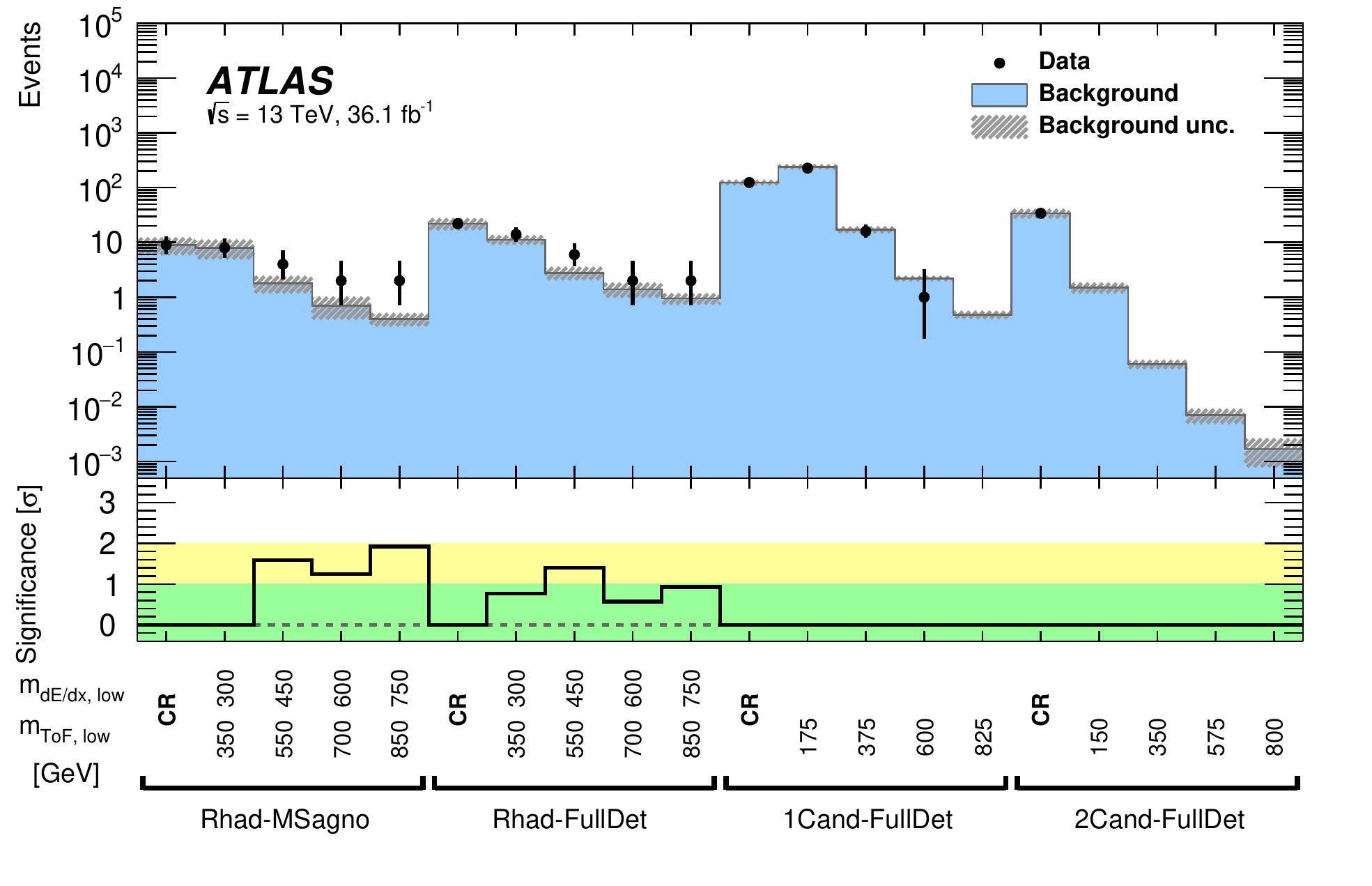}
  \caption{Expected and observed events in the 16 discovery regions introduced in Section~\ref{sec:selection} and for the respective CRs, including the systematic uncertainties. For data the Poisson error is shown. The lower plot shows the significance for SRs with a surplus of events between observed and estimated. The bin contents of SRs with inclusive mass ranges are correlated.}
  \label{fig:discSummary}
\end{figure}

\begin{table}[tbp]
  \scriptsize
  \centering
  \caption{Expected signal yield ($N_\text{exp.}$) and acceptance ($a$) $\times$ efficiency ($\varepsilon$), estimated background ($N_\text{est.}$) and observed number of events in data ($N_\text{obs.}$) for the full range of simulated masses in the \msagno \rhad search.}
  \label{tab:finalnumbers_rhad_ma}
  \setlength\tabcolsep{1pt}
  \begin{tabular*}{\textwidth}{@{\extracolsep{\fill}}lD{.}{.}{3.0}D{,}{\pm}{-1}D{,}{\pm}{-1}D{,}{\pm}{-1}D{.}{.}{1.0}}
    \hline
    \hline
    & & \multicolumn{4}{c}{\srrhma (\selidcalo)} \\
    \cmidrule(lr){3-6}
    \rhad & \multicolumn{1}{c}{\splitcell{Simulated\\mass~[\GeV]}} & N_\text{exp.},\sigma_{N_\text{exp.}} & a\times\varepsilon,\sigma_{a\times\varepsilon} & N_\text{est.},\sigma_{N_\text{est.}} & \multicolumn{1}{c}{$N_\text{obs.}$} \\
    \hline
    \multirow{13}{*}{Gluino}              &  400 & 160000 , 30000 & 0.044 , 0.003 & 8.0 , 3.0 & 8 \\
                                          &  600 &  28000 ,  5000 & 0.086 , 0.004 & 3.0 , 1.0 & 7 \\
                                          &  800 &   6000 ,  1000 & 0.106 , 0.005 & 1.8 , 0.6 & 4 \\
                                          & 1000 &   1300 ,   200 & 0.114 , 0.005 & 1.0 , 0.3 & 2 \\
                                          & 1200 &    400 ,    70 & 0.129 , 0.006 & 0.7 , 0.3 & 2 \\
                                          & 1400 &    140 ,    30 & 0.148 , 0.007 & 0.6 , 0.2 & 2 \\
                                          & 1600 &     42 ,     7 & 0.143 , 0.007 & 0.5 , 0.2 & 2 \\
                                          & 1800 &     13 ,     2 & 0.134 , 0.007 & 0.4 , 0.1 & 2 \\
                                          & 2000 &    4.4 ,   0.8 & 0.126 , 0.006 & 0.4 , 0.1 & 2 \\
                                          & 2200 &    1.5 ,   0.3 & 0.114 , 0.004 & 0.4 , 0.1 & 2 \\
                                          & 2400 &   0.51 ,  0.09 & 0.106 , 0.004 & 0.4 , 0.1 & 2 \\
                                          & 2600 &   0.18 ,  0.03 & 0.101 , 0.004 & 0.4 , 0.1 & 2 \\
                                          & 2800 &   0.06 ,  0.01 & 0.090 , 0.004 & 0.4 , 0.1 & 2 \\
                                          & 3000 &  0.023 , 0.004 & 0.090 , 0.004 & 0.4 , 0.1 & 2 \\
    \hline
    \multirow{5}{*}{\shortstack{Sbottom}} &  600 &    400 ,    80 & 0.063 , 0.003 & 3.0 , 1.0 & 7 \\
                                          &  800 &     80 ,    20 & 0.083 , 0.004 & 1.8 , 0.6 & 4 \\
                                          & 1000 &     19 ,     3 & 0.087 , 0.004 & 1.0 , 0.3 & 2 \\
                                          & 1200 &    5.4 ,   0.9 & 0.093 , 0.004 & 0.7 , 0.3 & 2 \\
                                          & 1400 &    1.5 ,   0.3 & 0.093 , 0.004 & 0.6 , 0.2 & 2 \\
    \hline
    \multirow{5}{*}{\shortstack{Stop}}    &  600 &    600 ,   100 & 0.095 , 0.005 & 3.0 , 1.0 & 7 \\
                                          &  800 &    120 ,   200 & 0.117 , 0.005 & 1.8 , 0.6 & 4 \\
                                          & 1000 &     28 ,     5 & 0.128 , 0.005 & 1.0 , 0.3 & 2 \\
                                          & 1200 &      8 ,     1 & 0.139 , 0.005 & 0.7 , 0.3 & 2 \\
                                          & 1400 &    2.4 ,   0.4 & 0.146 , 0.005 & 0.6 , 0.2 & 2 \\
    \hline
    \hline
  \end{tabular*}
\end{table}

\begin{table}[tbp]
  \scriptsize
  \centering
  \caption{Expected signal yield ($N_\text{exp.}$) and acceptance ($a$) $\times$ efficiency ($\varepsilon$), estimated background ($N_\text{est.}$) and observed number of events in data ($N_\text{obs.}$) for the full range of simulated masses in the full-detector \rhad search.}
  \label{tab:finalnumbers_rhad_fd}
  \setlength\tabcolsep{1pt}
  \begin{tabular*}{\textwidth}{@{\extracolsep{\fill}}lD{.}{.}{3.0}D{,}{\pm}{-1}D{,}{\pm}{-1}D{,}{\pm}{-1}D{.}{.}{1.0}D{,}{\pm}{-1}D{,}{\pm}{-1}D{,}{\pm}{-1}D{.}{.}{1.0}}
    \hline
    \hline
    & & \multicolumn{4}{c}{\srrhfd (\selloose)} & \multicolumn{4}{c}{\srrhfd (\selidcalo)} \\
    \cmidrule(lr){3-6}\cmidrule(lr){7-10}
    \rhad & \multicolumn{1}{c}{\splitcell{Simulated\\mass~[\GeV]}} & N_\text{exp.},\sigma_{N_\text{exp.}} & a\times\varepsilon,\sigma_{a\times\varepsilon} & N_\text{est.},\sigma_{N_\text{est.}} & \multicolumn{1}{c}{$N_\text{obs.}$} & N_\text{exp.},\sigma_{N_\text{exp.}} & a\times\varepsilon,\sigma_{a\times\varepsilon} & N_\text{est.},\sigma_{N_\text{est.}} & \multicolumn{1}{c}{$N_\text{obs.}$} \\
    \hline
    \multirow{13}{*}{Gluino}              &  400 & 60000     , 20000     & 0.016 , 0.002 &  1.5 ,  0.5 & 1 & 160000     , 3000     & 0.044 , 0.003 & 9.0 , 2.0& 13 \\
                                          &  600 & 11000     ,  4000     & 0.033 , 0.003 &  0.5 ,  0.2 & 1 &  24000     , 4000     & 0.071 , 0.004 & 4.0 , 1.0 &  9 \\
                                          &  800 &  2400     ,   600     & 0.044 , 0.003 &  0.3 ,  0.1 & 1 &   4500     ,  800     & 0.083 , 0.004 & 2.5 , 0.7 &  5 \\
                                          & 1000 &   500     ,   100     & 0.045 , 0.003 & 0.14 , 0.05 & 0 &   1100     ,  200     & 0.091 , 0.005 & 1.6 , 0.4 &  3 \\
                                          & 1200 &   160     ,    40     & 0.053 , 0.004 & 0.10 , 0.04 & 0 &    300     ,   50     & 0.096 , 0.005 & 1.3 , 0.4 &  2 \\
                                          & 1400 &    60     ,    10     & 0.063 , 0.005 & 0.07 , 0.03 & 0 &    100     ,   20     & 0.104 , 0.006 & 1.1 , 0.3 &  2 \\
                                          & 1600 &    17     ,     4     & 0.057 , 0.004 & 0.06 , 0.03 & 0 &     30     ,    6     & 0.104 , 0.006 & 1.0 , 0.3 &  2 \\
                                          & 1800 &     5     ,     1     & 0.052 , 0.004 & 0.05 , 0.03 & 0 &     10     ,    2     & 0.099 , 0.006 & 0.9 , 0.3 &  2 \\
                                          & 2000 &     1.9   ,     0.4   & 0.053 , 0.003 & 0.05 , 0.02 & 0 &      2.9   ,    0.6   & 0.083 , 0.004 & 0.9 , 0.2 &  2 \\
                                          & 2200 &     0.6   ,     0.1   & 0.043 , 0.003 & 0.05 , 0.02 & 0 &      1.0   ,    0.2   & 0.079 , 0.003 & 0.9 , 0.2 &  2 \\
                                          & 2400 &     0.18  ,     0.04  & 0.037 , 0.002 & 0.05 , 0.02 & 0 &      0.38  ,    0.07  & 0.079 , 0.004 & 0.9 , 0.2 &  2 \\
                                          & 2600 &     0.07  ,     0.01  & 0.036 , 0.002 & 0.05 , 0.02 & 0 &      0.13  ,    0.02  & 0.074 , 0.003 & 0.9 , 0.2 &  2 \\
                                          & 2800 &     0.019 ,     0.004 & 0.027 , 0.002 & 0.05 , 0.02 & 0 &      0.049 ,    0.009 & 0.071 , 0.003 & 0.9 , 0.2 &  2 \\
                                          & 3000 &     0.007 ,     0.002 & 0.028 , 0.002 & 0.05 , 0.02 & 0 &      0.017 ,    0.003 & 0.066 , 0.003 & 0.9 , 0.2 &  2 \\
    \hline
    \multirow{5}{*}{\shortstack{Sbottom}} &  600 &   200     ,    50     & 0.032 , 0.002 &  0.5 ,  0.2 & 1 &    300     ,   60     & 0.047 , 0.003 & 4.0 , 1.0 &  9 \\
                                          &  800 &    38     ,     8     & 0.037 , 0.003 &  0.3 ,  0.1 & 1 &     60     ,   10     & 0.061 , 0.003 & 2.5 , 0.7 &  5 \\
                                          & 1000 &     9     ,     2     & 0.040 , 0.003 & 0.14 , 0.05 & 0 &     14     ,    3     & 0.064 , 0.003 & 1.6 , 0.4 &  3 \\
                                          & 1200 &     2.5   ,     0.5   & 0.043 , 0.003 & 0.10 , 0.04 & 0 &      3.9   ,    0.7   & 0.068 , 0.003 & 1.3 , 0.4 &  2 \\
                                          & 1400 &     0.7   ,     0.1   & 0.042 , 0.003 & 0.07 , 0.03 & 0 &      1.1   ,    0.2   & 0.069 , 0.003 & 1.1 , 0.3 &  2 \\
    \hline
    \multirow{5}{*}{\shortstack{Stop}}    &  600 &   390     ,   70      & 0.062 , 0.004 &  0.5 ,  0.2 & 1 &    370     ,   70     & 0.059 , 0.004 &   4 , 1   &  9 \\
                                          &  800 &    80     ,   20      & 0.075 , 0.004 &  0.3 ,  0.1 & 1 &     80     ,   20     & 0.077 , 0.004 & 2.5 , 0.7 &  5 \\
                                          & 1000 &    18     ,    4      & 0.083 , 0.004 & 0.14 , 0.05 & 0 &     18     ,    3     & 0.081 , 0.004 & 1.6 , 0.4 &  3 \\
                                          & 1200 &     5     ,    1      & 0.088 , 0.004 & 0.10 , 0.04 & 0 &      4.9   ,    0.9   & 0.085 , 0.004 & 1.3 , 0.4 &  2 \\
                                          & 1400 &     1.6   ,    0.3    & 0.093 , 0.005 & 0.07 , 0.03 & 0 &      1.5   ,    0.3   & 0.089 , 0.004 & 1.1 , 0.3 &  2 \\
    \hline
    \hline
  \end{tabular*}
\end{table}

\begin{table}[tbp]
  \footnotesize
  \centering
  \caption{\pzero-values and model-independent upper limits on cross section ($\sigma$) $\times$ acceptance ($a$) $\times$ efficiency ($\varepsilon$) for the 16 discovery regions.}
  \label{tab:pzeros}
  \setlength\tabcolsep{1pt}
  \begin{tabular*}{\textwidth}{@{\extracolsep{\fill}}lD{.}{.}{3.0}D{.}{.}{3.0}D{,}{\pm}{-1}D{.}{.}{3.0}D{.}{.}{1.3}D{.}{.}{1.2}D{.}{.}{1.2}}
    \hline
    \hline
    & \multicolumn{2}{c}{Lower mass requirements} & & & & & \\
    \cmidrule(lr){2-3}
    Selection & \multicolumn{1}{c}{\mtofmin [\GeV]} & \multicolumn{1}{c}{\mdedxmin [\GeV]} & N_\text{est.},\sigma_{N_\text{est.}} & \multicolumn{1}{c}{$N_\text{obs.}$} & \multicolumn{1}{c}{\pzero} & \multicolumn{1}{c}{Significance [$\sigma$]} & \multicolumn{1}{c}{\splitcell{95\% CL upper limit\\$\sigma\times{a}\times\varepsilon$ [fb]}} \\
    \hline
    \multirow{4}{*}{\srrhma}  & 350 & 300 &   8.0    ,  3.0    &   8 & 0.5   &      & 0.25 \\
                              & 550 & 450 &   1.8    ,  0.6    &   4 & 0.056 & 1.59 & 0.20 \\
                              & 700 & 600 &   0.7    ,  0.3    &   2 & 0.11  & 1.24 & 0.17 \\
                              & 850 & 750 &   0.4    ,  0.1    &   2 & 0.028 & 1.92 & 0.17 \\
    \hline
    \multirow{4}{*}{\srrhfd}  & 350 & 300 &  11      ,  2      &  14 & 0.22  & 0.77 & 0.42 \\
                              & 550 & 450 &   2.8    ,  0.7    &   6 & 0.081 & 1.40 & 0.25 \\
                              & 700 & 600 &   1.4    ,  0.4    &   2 & 0.28  & 0.57 & 0.14 \\
                              & 850 & 750 &   0.95   ,  0.2    &   2 & 0.18  & 0.93 & 0.14 \\
    \hline
    \multirow{4}{*}{\srone}   & 175 &     & 240      , 20      & 227 & 0.5   &      & 1.26 \\
                              & 375 &     &  17      ,  2      &  16 & 0.5   &      & 0.24 \\
                              & 600 &     &   2.2    ,  0.2    &   1 & 0.5   &      & 0.10 \\
                              & 825 &     &   0.48   ,  0.07   &   0 & 0.5   &      & 0.08 \\
    \hline
    \multirow{4}{*}{\srtwo}   & 150 &     &   1.5    ,  0.3    &   0 & 0.5   &      & 0.09 \\
                              & 350 &     &   0.06   ,  0.01   &   0 & 0.5   &      & 0.08 \\
                              & 575 &     &   0.007  ,  0.002  &   0 & 0.5   &      & 0.08 \\
                              & 800 &     &   0.0017 ,  0.0009 &   0 & 0.5   &      & 0.08 \\
    \hline
    \hline
  \end{tabular*}
\end{table}

\begin{table}[tbp]
  \scriptsize
  \centering
  \caption{Expected signal yield ($N_\text{exp.}$) and acceptance ($a$) $\times$ efficiency ($\varepsilon$), estimated background ($N_\text{est.}$) and observed number of events in data ($N_\text{obs.}$) for the full range of simulated masses in the \msagno search for metastable gluino \rhads.}
  \label{tab:finalnumbers_rhad_ma_meta}
  \setlength\tabcolsep{1pt}
  \begin{tabular*}{\textwidth}{@{\extracolsep{\fill}}D{.}{.}{4.0}D{,}{\pm}{-1}D{,}{\pm}{-1}D{,}{\pm}{-1}D{,}{\pm}{-1}D{,}{\pm}{-1}D{,}{\pm}{-1}D{,}{\pm}{-1}D{.}{.}{1.0}}
    \hline
    \hline
    \multicolumn{1}{c}{Gluino \rhad} & \multicolumn{8}{c}{\srrhma (\selidcalo)} \\
    \cmidrule(lr){2-9}
    \multicolumn{1}{c}{Lifetime} & \multicolumn{2}{c}{10~ns} & \multicolumn{2}{c}{30~ns} & \multicolumn{2}{c}{50~ns} & \multicolumn{2}{c}{} \\
    \cmidrule(lr){2-3}\cmidrule(lr){4-5}\cmidrule(lr){6-7}
    \multicolumn{1}{c}{\splitcell{Simulated\\mass~[\GeV]}} & N_\text{exp.},\sigma_{N_\text{exp.}} & a\times\varepsilon,\sigma_{a\times\varepsilon} & N_\text{exp.},\sigma_{N_\text{exp.}} & a\times\varepsilon,\sigma_{a\times\varepsilon} & N_\text{exp.},\sigma_{N_\text{exp.}} & a\times\varepsilon,\sigma_{a\times\varepsilon} & N_\text{est.},\sigma_{N_\text{est.}} & \multicolumn{1}{c}{$N_\text{obs.}$} \\
    \hline
    1000 &  800 ,  100 & 0.065 , 0.004 & 1400 ,  300 & 0.121 , 0.006 & 1500 , 300 & 0.125 , 0.005 & 1.0 , 0.3 & 2 \\
    1200 &  220 ,   40 & 0.072 , 0.004 &  400 ,   70 & 0.129 , 0.006 &  410 ,  70 & 0.133 , 0.005 & 0.7 , 0.3 & 2 \\
    1400 &   70 ,   10 & 0.079 , 0.004 &  120 ,   20 & 0.132 , 0.005 &  140 ,  30 & 0.149 , 0.006 & 0.6 , 0.2 & 2 \\
    1600 &   22 ,    4 & 0.074 , 0.003 &   41 ,    7 & 0.140 , 0.005 &   41 ,   7 & 0.142 , 0.005 & 0.5 , 0.2 & 2 \\
    1800 &    8 ,    1 & 0.077 , 0.003 &   14 ,    2 & 0.139 , 0.005 &   14 ,   2 & 0.142 , 0.005 & 0.4 , 0.1 & 2 \\
    2000 &  2.8 ,  0.5 & 0.080 , 0.005 &  4.7 ,  0.8 & 0.132 , 0.007 &  5.2 , 0.9 & 0.146 , 0.005 & 0.4 , 0.1 & 2 \\
    2200 &  1.0 ,  0.2 & 0.075 , 0.004 &  1.7 ,  0.3 & 0.132 , 0.005 &  1.7 , 0.3 & 0.130 , 0.004 & 0.4 , 0.1 & 2 \\
    2400 & 0.35 , 0.06 & 0.073 , 0.004 & 0.58 , 0.10 & 0.120 , 0.004 &  0.6 , 0.1 & 0.122 , 0.004 & 0.4 , 0.1 & 2 \\
    \hline
    \hline
  \end{tabular*}
\end{table}

\begin{table}[tbp]
  \scriptsize
  \centering
  \caption{Expected signal yield ($N_\text{exp.}$) and acceptance ($a$) $\times$ efficiency ($\varepsilon$), estimated background ($N_\text{est.}$) and observed number of events in data ($N_\text{obs.}$) for the full range of simulated masses in the full-detector direct-stau search.}
  \label{tab:finalnumbers_stau}
  \setlength\tabcolsep{1pt}
  \begin{tabular*}{\textwidth}{@{\extracolsep{\fill}}D{.}{.}{3.0}D{,}{\pm}{-1}D{,}{\pm}{-1}D{,}{\pm}{-1}D{.}{.}{1.0}D{,}{\pm}{-1}D{,}{\pm}{-1}D{,}{\pm}{-1}D{.}{.}{2.0}}
    \hline
    \hline
    & \multicolumn{4}{c}{\srtwo} & \multicolumn{4}{c}{\srone} \\
    \cmidrule(lr){2-5}\cmidrule(lr){6-9}
    \multicolumn{1}{c}{\splitcell{Simulated\\mass~[\GeV]}} & N_\text{exp.},\sigma_{N_\text{exp.}} & a\times\varepsilon,\sigma_{a\times\varepsilon} & N_\text{est.},\sigma_{N_\text{est.}} & \multicolumn{1}{c}{$N_\text{obs.}$} & N_\text{exp.},\sigma_{N_\text{exp.}} & a\times\varepsilon,\sigma_{a\times\varepsilon} & N_\text{est.},\sigma_{N_\text{est.}} & \multicolumn{1}{c}{$N_\text{obs.}$} \\
    \hline
    287 & 13     , 1     & 0.167 , 0.005 &  0.33  , 0.06  & 0 & 5.1   , 0.6   & 0.068 , 0.003 & 80.0 , 7.0 & 74 \\
    318 &  9     , 1     & 0.179 , 0.007 &  0.22  , 0.04  & 0 & 3.6   , 0.4   & 0.073 , 0.004 & 56.0 , 5.0 & 52 \\
    349 &  6.1   , 0.7   & 0.181 , 0.005 &  0.15  , 0.03  & 0 & 2.5   , 0.2   & 0.076 , 0.003 & 41.0 , 4.0 & 36 \\
    380 &  4.3   , 0.6   & 0.184 , 0.006 &  0.11  , 0.02  & 0 & 2.1   , 0.2   & 0.089 , 0.005 & 30.0 , 3.0 & 24 \\
    411 &  3.2   , 0.4   & 0.196 , 0.005 &  0.08  , 0.02  & 0 & 1.5   , 0.1   & 0.093 , 0.004 & 23.0 , 2.0 & 20 \\
    442 &  2.4   , 0.3   & 0.198 , 0.007 &  0.06  , 0.01  & 0 & 1.2   , 0.2   & 0.096 , 0.005 & 17.0 , 2.0 & 16 \\
    473 &  1.8   , 0.3   & 0.204 , 0.005 &  0.045 , 0.009 & 0 & 0.92  , 0.09  & 0.105 , 0.004 & 13.0 , 1.0 & 15 \\
    504 &  1.4   , 0.2   & 0.210 , 0.005 &  0.035 , 0.007 & 0 & 0.68  , 0.06  & 0.105 , 0.004 & 10.1 , 1.0 & 11 \\
    536 &  1.0   , 0.1   & 0.208 , 0.005 &  0.027 , 0.006 & 0 & 0.55  , 0.06  & 0.111 , 0.004 &  7.9 , 0.8 &  7 \\
    567 &  0.84  , 0.10  & 0.224 , 0.006 &  0.027 , 0.006 & 0 & 0.43  , 0.04  & 0.113 , 0.004 &  6.3 , 0.6 &  4 \\
    598 &  0.65  , 0.09  & 0.227 , 0.006 &  0.022 , 0.005 & 0 & 0.34  , 0.03  & 0.118 , 0.004 &  5.0 , 0.5 &  3 \\
    629 &  0.50  , 0.07  & 0.227 , 0.006 &  0.017 , 0.004 & 0 & 0.27  , 0.02  & 0.124 , 0.004 &  5.0 , 0.5 &  3 \\
    660 &  0.40  , 0.05  & 0.234 , 0.006 &  0.014 , 0.003 & 0 & 0.22  , 0.02  & 0.125 , 0.005 &  4.0 , 0.4 &  3 \\
    692 &  0.30  , 0.05  & 0.224 , 0.008 &  0.011 , 0.003 & 0 & 0.17  , 0.02  & 0.125 , 0.005 &  3.2 , 0.3 &  2 \\
    723 &  0.24  , 0.03  & 0.229 , 0.007 &  0.009 , 0.002 & 0 & 0.13  , 0.01  & 0.120 , 0.005 &  2.6 , 0.3 &  1 \\
    754 &  0.19  , 0.02  & 0.224 , 0.006 &  0.008 , 0.002 & 0 & 0.112 , 0.009 & 0.132 , 0.004 &  2.2 , 0.2 &  1 \\
    785 &  0.15  , 0.02  & 0.222 , 0.006 &  0.007 , 0.002 & 0 & 0.091 , 0.007 & 0.135 , 0.005 &  1.8 , 0.2 &  0 \\
    817 &  0.12  , 0.01  & 0.219 , 0.006 &  0.007 , 0.002 & 0 & 0.073 , 0.006 & 0.134 , 0.004 &  1.5 , 0.1 &  0 \\
    848 &  0.09  , 0.01  & 0.215 , 0.005 &  0.006 , 0.001 & 0 & 0.061 , 0.005 & 0.138 , 0.004 &  1.3 , 0.1 &  0 \\
    879 &  0.08  , 0.01  & 0.212 , 0.005 &  0.005 , 0.001 & 0 & 0.052 , 0.005 & 0.146 , 0.005 &  1.3 , 0.1 &  0 \\
    911 &  0.065 , 0.007 & 0.225 , 0.006 &  0.004 , 0.001 & 0 & 0.041 , 0.003 & 0.144 , 0.005 &  1.1 , 0.1 &  0 \\
    \hline
    \hline
  \end{tabular*}
\end{table}

\begin{table}[tbp]
  \scriptsize
  \centering
  \caption{Expected signal yield ($N_\text{exp.}$) and acceptance ($a$) $\times$ efficiency ($\varepsilon$), estimated background ($N_\text{est.}$) and observed number of events in data ($N_\text{obs.}$) for the full range of simulated masses in the full-detector chargino search.}
  \label{tab:finalnumbers_chargino}
  \setlength\tabcolsep{1pt}
  \begin{tabular*}{\textwidth}{@{\extracolsep{\fill}}D{.}{.}{4.0}D{,}{\pm}{-1}D{,}{\pm}{-1}D{,}{\pm}{-1}D{.}{.}{1.0}D{,}{\pm}{-1}D{,}{\pm}{-1}D{,}{\pm}{-1}D{.}{.}{3.0}}
    \hline
    \hline
    & \multicolumn{4}{c}{\srtwo} & \multicolumn{4}{c}{\srone} \\
    \cmidrule(lr){2-5}\cmidrule(lr){6-9}
    \multicolumn{1}{c}{\splitcell{Simulated\\mass~[\GeV]}} & N_\text{exp.},\sigma_{N_\text{exp.}} & a\times\varepsilon,\sigma_{a\times\varepsilon} & N_\text{est.},\sigma_{N_\text{est.}} & \multicolumn{1}{c}{$N_\text{obs.}$} & N_\text{exp.},\sigma_{N_\text{exp.}} & a\times\varepsilon,\sigma_{a\times\varepsilon} & N_\text{est.},\sigma_{N_\text{est.}} & \multicolumn{1}{c}{$N_\text{obs.}$} \\
    \hline
     200 & 2600   , 400   & 0.083 , 0.003 &    1.5 ,   0.3  & 0 & 1200 , 200   & 0.038 , 0.002 &  230 ,   20 & 227 \\
     250 & 1200   , 200   & 0.091 , 0.003 &   0.51 ,  0.10  & 0 &  800 , 100   & 0.062 , 0.003 &  110 ,   10 & 109 \\
     300 &  690   , 100   & 0.102 , 0.004 &   0.33 ,  0.06  & 0 &  490 ,  50   & 0.073 , 0.003 &   79 ,    7 &  74 \\
     350 &  360   ,  50   & 0.101 , 0.004 &   0.15 ,  0.03  & 0 &  280 ,  30   & 0.078 , 0.003 &   41 ,    4 &  36 \\
     400 &  220   ,  30   & 0.107 , 0.004 &   0.08 ,  0.02  & 0 &  180 ,  20   & 0.089 , 0.004 &   23 ,    2 &  20 \\
     450 &  140   ,  20   & 0.113 , 0.004 &   0.06 ,  0.01  & 0 &  120 ,  10   & 0.100 , 0.004 &   17 ,    2 &  16 \\
     500 &   90   ,  10   & 0.115 , 0.004 &  0.034 , 0.007  & 0 &   77 ,   8   & 0.100 , 0.004 &   10 ,    1 &  11 \\
     550 &   59   ,   8   & 0.119 , 0.004 &  0.027 , 0.006  & 0 &   52 ,   5   & 0.105 , 0.004 &  7.9 ,  0.8 &   7 \\
     600 &   42   ,   6   & 0.129 , 0.004 &  0.021 , 0.004  & 0 &   36 ,   4   & 0.110 , 0.004 &  5.0 ,  0.5 &   3 \\
     650 &   27   ,   4   & 0.123 , 0.004 &  0.014 , 0.003  & 0 &   24 ,   2   & 0.107 , 0.004 &  4.0 ,  0.4 &   3 \\
     700 &   18   ,   3   & 0.122 , 0.004 &  0.011 , 0.003  & 0 &   17 ,   2   & 0.113 , 0.004 &  3.2 ,  0.3 &   2 \\
     750 &   12   ,   2   & 0.113 , 0.004 &  0.008 , 0.002  & 0 &   13 ,   1   & 0.118 , 0.004 &  2.1 ,  0.2 &   1 \\
     800 &    9   ,   1   & 0.120 , 0.004 &  0.007 , 0.002  & 0 &  9.2 ,   0.9 & 0.123 , 0.004 &  1.8 ,  0.2 &   0 \\
     850 &    6.0 ,   0.8 & 0.112 , 0.005 &  0.006 , 0.001  & 0 &  6.1 ,   0.6 & 0.114 , 0.005 &  1.3 ,  0.1 &   0 \\
     900 &    4.2 ,   0.6 & 0.108 , 0.004 &  0.004 , 0.001  & 0 &  4.7 ,   0.5 & 0.121 , 0.004 &  1.1 ,  0.1 &   0 \\
     950 &    3.2 ,   0.5 & 0.112 , 0.004 &  0.003 , 0.001  & 0 &  3.3 ,   0.3 & 0.118 , 0.004 &  1.0 ,  0.1 &   0 \\
    1000 &    2.2 ,   0.4 & 0.106 , 0.005 & 0.0029 , 0.0009 & 0 &  2.5 ,   0.2 & 0.120 , 0.006 & 0.84 , 0.10 &   0 \\
    1100 &    1.2 ,   0.2 & 0.105 , 0.004 & 0.0019 , 0.0007 & 0 &  1.5 ,   0.2 & 0.131 , 0.004 & 0.54 , 0.07 &   0 \\
    1200 &   0.62 ,  0.09 & 0.096 , 0.004 & 0.0015 , 0.0006 & 0 & 0.74 ,  0.07 & 0.115 , 0.004 & 0.42 , 0.06 &   0 \\
    1300 &   0.32 ,  0.04 & 0.087 , 0.003 & 0.0012 , 0.0006 & 0 & 0.44 ,  0.05 & 0.118 , 0.004 & 0.33 , 0.05 &   0 \\
    1400 &   0.19 ,  0.03 & 0.087 , 0.004 & 0.0009 , 0.0005 & 0 & 0.26 ,  0.03 & 0.120 , 0.004 & 0.27 , 0.04 &   0 \\
    1500 &   0.10 ,  0.02 & 0.077 , 0.003 & 0.0007 , 0.0005 & 0 & 0.16 ,  0.01 & 0.121 , 0.004 & 0.21 , 0.04 &   0 \\
    \hline
    \hline
  \end{tabular*}
\end{table}

\FloatBarrier

\begin{figure}[tbp]
  \centering
  \includegraphics[width=.48\textwidth]{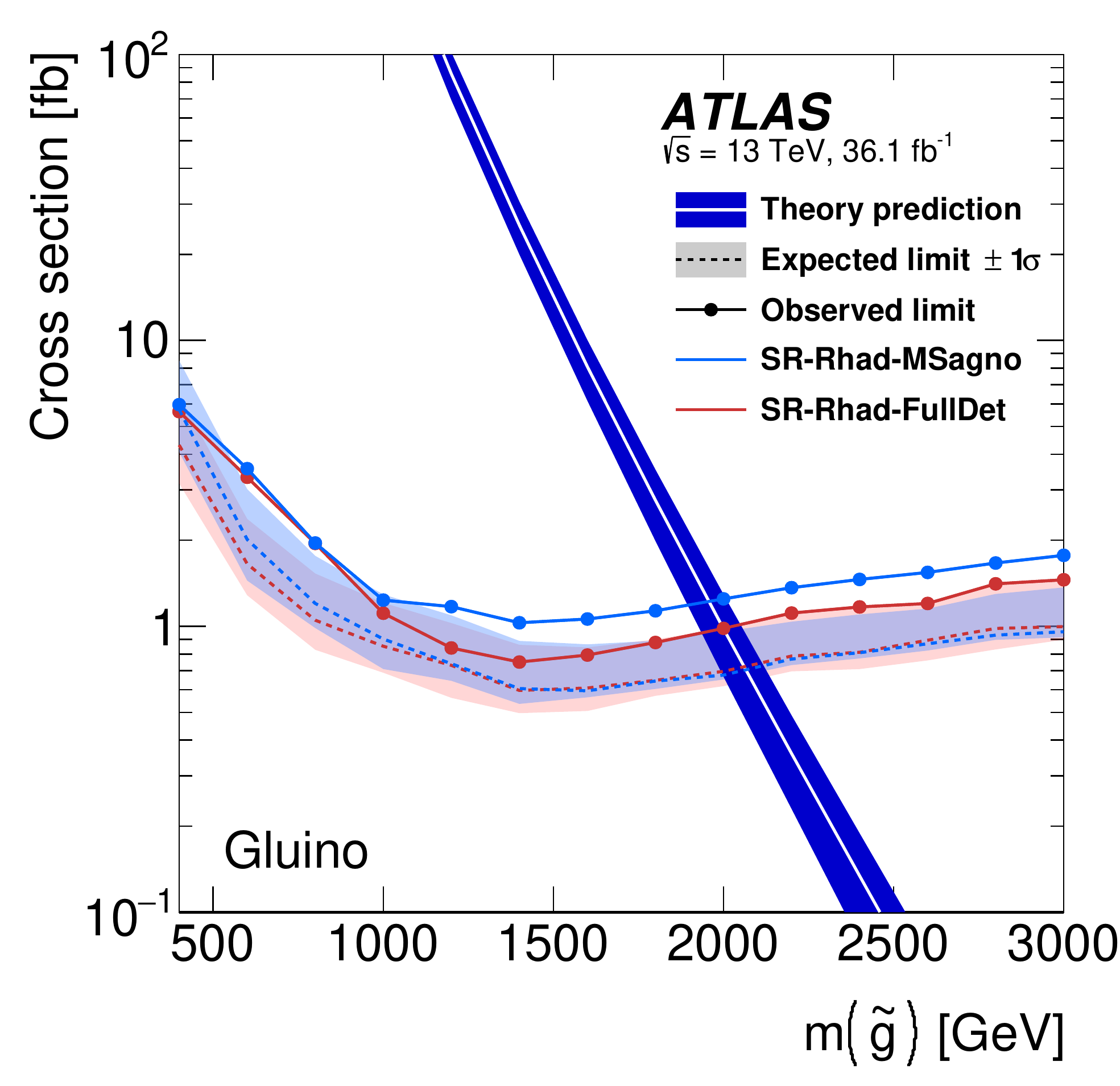}\\
  \includegraphics[width=.48\textwidth]{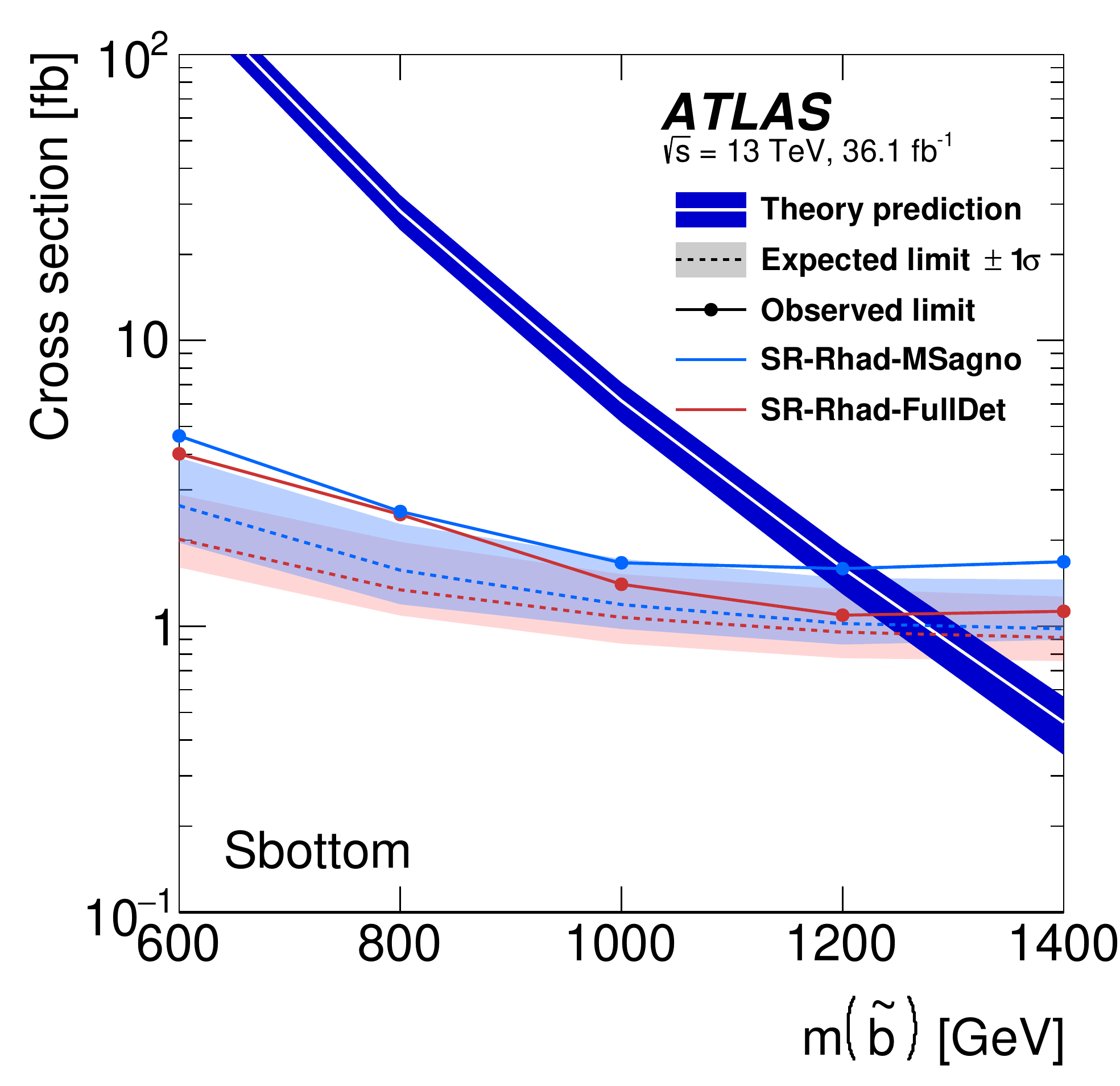}
  \includegraphics[width=.48\textwidth]{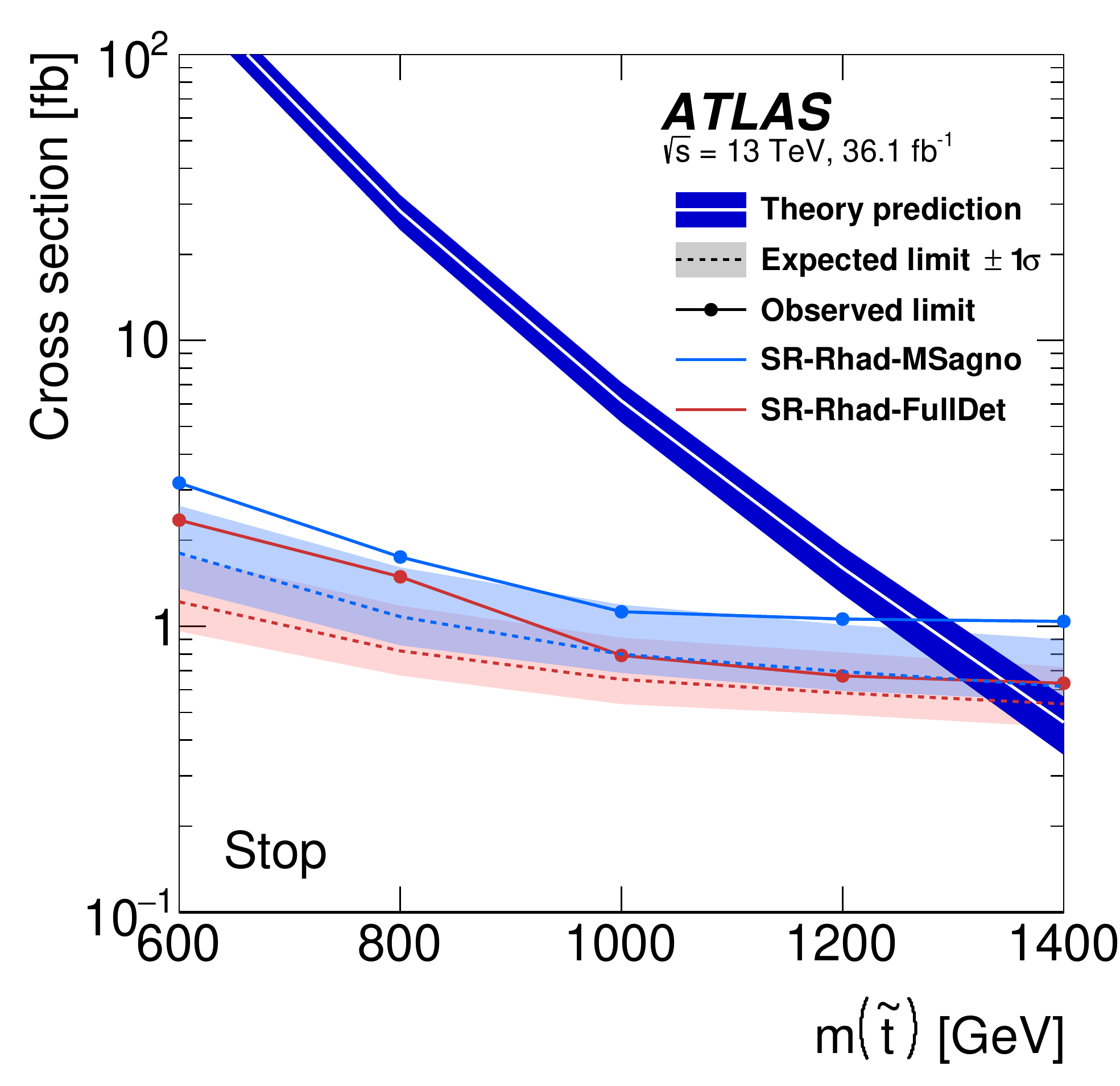}
  \caption{Expected (dashed lines) and observed (marked solid lines) upper cross-section limits in the gluino (top), sbottom (bottom-left) and stop (bottom-right) \rhad searches, respectively, using two independent and not to be combined approaches based on SRs \srrhma (light blue) and \srrhfd (red). The shaded light-blue/light-red bands represent the $\pm{1}\sigma$ uncertainties in the expected limits. The result obtained using the former SR has a much reduced dependence on the modeling of \rhad interaction with matter with respect to the other, therefore the two results must not combined. The theory prediction along with its $\pm1\sigma$ uncertainty is shown as a white line and a dark-blue band, respectively.}
  \label{fig:limitrhadComb}
\end{figure}

\begin{figure}[tbp]
  \centering
  \includegraphics[width=.48\textwidth]{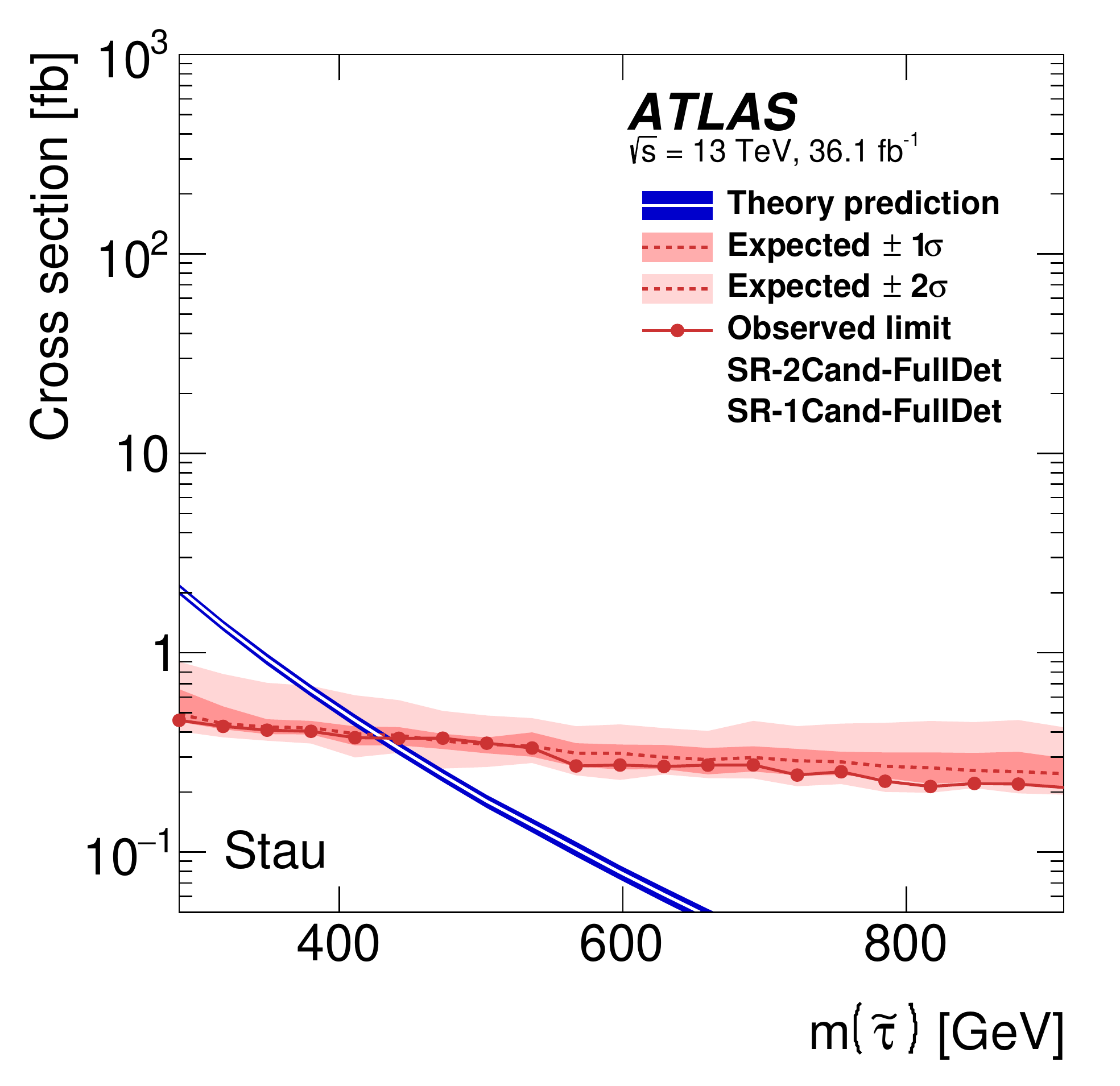}
  \includegraphics[width=.48\textwidth]{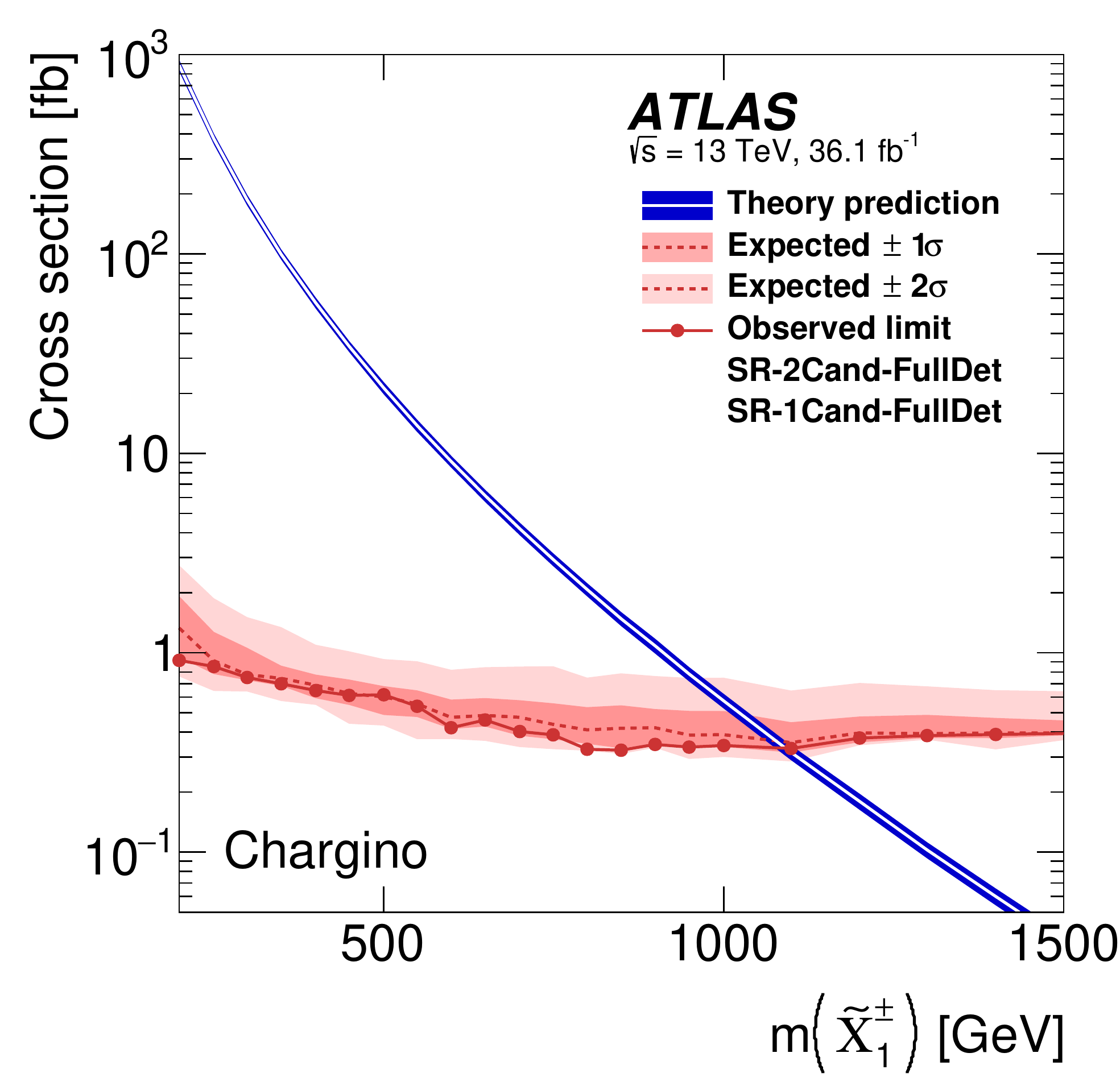}
  \caption{Expected (dashed red line) and observed (marked solid red line) upper cross-section limits using combined two- and one-candidate SRs for stau pair production (left) and chargino pair production (right). The shaded dark-red (light-red) bands represent the $\pm{1}\sigma$ ($\pm{2}\sigma$) uncertainties in the expected limits. The theory prediction along with its $\pm1\sigma$ uncertainty is shown as a white line and a blue band, respectively.}
  \label{fig:limit2Cand1Cand}
\end{figure}

\begin{figure}[tbp]
  \centering
  \includegraphics[width=.48\textwidth]{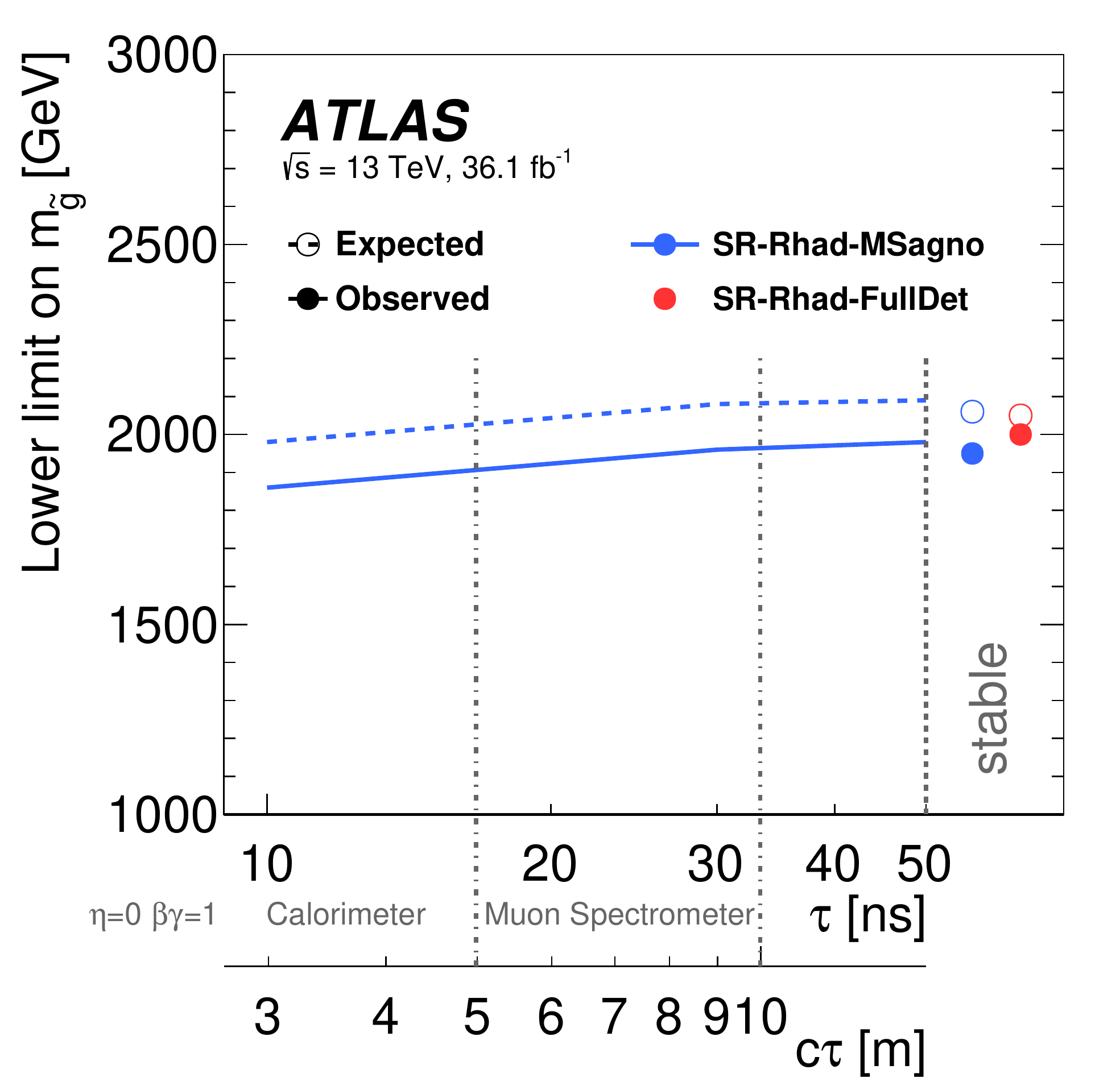}
  \caption{Expected (dashed blue line) and observed (solid blue line) lower mass limits at the 95\% CL level for gluino \rhads with different mean lifetimes derived using the \msagno analysis (\srrhma). Expected (empty circles) and observed (filled circles) limits on stable gluino \rhads are shown for both the \msagno (\srrhma) and the full-detector analysis (\srrhfd).}
  \label{fig:limitsRhadGluinoLifetimeSummary}
\end{figure}

\FloatBarrier
\section{Conclusion}
\label{sec:conclusion}

A search for heavy, charged, long-lived particles is performed using a data sample of \lumi of proton--proton collisions at $\sqrt{s} = 13~\TeV$ collected by the ATLAS experiment at the Large Hadron Collider at CERN. The search is based on observables related to large ionization losses, measured in the innermost tracking detector, and slow propagation velocities, measured in the tile calorimeter and muon spectrometer. Both observables are signatures of heavy charged particles traveling significantly slower than the speed of light. No significant deviations from the expected background are observed. Upper limits at 95\% confidence level are provided on the production cross sections of long-lived \rhads, as well as directly pair-produced staus and charginos. These results translate into lower limits on the masses of long-lived gluino, sbottom and stop \rhads, as well as staus and charginos of \limRHFDgluObs, \limRHFDsboObs, \limRHFDstoObs, \limStauObs and \limCharObs, respectively.

\section*{Acknowledgments}


We thank CERN for the very successful operation of the LHC, as well as the
support staff from our institutions without whom ATLAS could not be
operated efficiently.

We acknowledge the support of ANPCyT, Argentina; YerPhI, Armenia; ARC, Australia; BMWFW and FWF, Austria; ANAS, Azerbaijan; SSTC, Belarus; CNPq and FAPESP, Brazil; NSERC, NRC and CFI, Canada; CERN; CONICYT, Chile; CAS, MOST and NSFC, China; COLCIENCIAS, Colombia; MSMT CR, MPO CR and VSC CR, Czech Republic; DNRF and DNSRC, Denmark; IN2P3-CNRS, CEA-DRF/IRFU, France; SRNSFG, Georgia; BMBF, HGF, and MPG, Germany; GSRT, Greece; RGC, Hong Kong SAR, China; ISF and Benoziyo Center, Israel; INFN, Italy; MEXT and JSPS, Japan; CNRST, Morocco; NWO, Netherlands; RCN, Norway; MNiSW and NCN, Poland; FCT, Portugal; MNE/IFA, Romania; MES of Russia and NRC KI, Russian Federation; JINR; MESTD, Serbia; MSSR, Slovakia; ARRS and MIZ\v{S}, Slovenia; DST/NRF, South Africa; MINECO, Spain; SRC and Wallenberg Foundation, Sweden; SERI, SNSF and Cantons of Bern and Geneva, Switzerland; MOST, Taiwan; TAEK, Turkey; STFC, United Kingdom; DOE and NSF, United States of America. In addition, individual groups and members have received support from BCKDF, CANARIE, CRC and Compute Canada, Canada; COST, ERC, ERDF, Horizon 2020, and Marie Sk{\l}odowska-Curie Actions, European Union; Investissements d' Avenir Labex and Idex, ANR, France; DFG and AvH Foundation, Germany; Herakleitos, Thales and Aristeia programmes co-financed by EU-ESF and the Greek NSRF, Greece; BSF-NSF and GIF, Israel; CERCA Programme Generalitat de Catalunya, Spain; The Royal Society and Leverhulme Trust, United Kingdom.

The crucial computing support from all WLCG partners is acknowledged gratefully, in particular from CERN, the ATLAS Tier-1 facilities at TRIUMF (Canada), NDGF (Denmark, Norway, Sweden), CC-IN2P3 (France), KIT/GridKA (Germany), INFN-CNAF (Italy), NL-T1 (Netherlands), PIC (Spain), ASGC (Taiwan), RAL (UK) and BNL (USA), the Tier-2 facilities worldwide and large non-WLCG resource providers. Major contributors of computing resources are listed in Ref.~\cite{ATL-GEN-PUB-2016-002}.

\printbibliography

\clearpage
 
\begin{flushleft}
{\Large The ATLAS Collaboration}

\bigskip

M.~Aaboud$^\textrm{\scriptsize 34d}$,    
G.~Aad$^\textrm{\scriptsize 99}$,    
B.~Abbott$^\textrm{\scriptsize 125}$,    
D.C.~Abbott$^\textrm{\scriptsize 100}$,    
O.~Abdinov$^\textrm{\scriptsize 13,*}$,    
B.~Abeloos$^\textrm{\scriptsize 129}$,    
D.K.~Abhayasinghe$^\textrm{\scriptsize 91}$,    
S.H.~Abidi$^\textrm{\scriptsize 164}$,    
O.S.~AbouZeid$^\textrm{\scriptsize 39}$,    
N.L.~Abraham$^\textrm{\scriptsize 153}$
% ...
\bigskip
\\

% ...
$^{13}$Institute of Physics, Azerbaijan Academy of Sciences, Baku; Azerbaijan.\\
% ...
$^{34}$$^{(a)}$Facult\'e des Sciences Ain Chock, R\'eseau Universitaire de Physique des Hautes Energies - Universit\'e Hassan II, Casablanca;$^{(b)}$Centre National de l'Energie des Sciences Techniques Nucleaires (CNESTEN), Rabat;$^{(c)}$Facult\'e des Sciences Semlalia, Universit\'e Cadi Ayyad, LPHEA-Marrakech;$^{(d)}$Facult\'e des Sciences, Universit\'e Mohamed Premier and LPTPM, Oujda;$^{(e)}$Facult\'e des sciences, Universit\'e Mohammed V, Rabat; Morocco.\\
% ...
$^{39}$Niels Bohr Institute, University of Copenhagen, Copenhagen; Denmark.\\
% ...
$^{91}$Department of Physics, Royal Holloway University of London, Egham; United Kingdom.\\
% ...
$^{99}$CPPM, Aix-Marseille Universit\'e, CNRS/IN2P3, Marseille; France.\\
$^{100}$Department of Physics, University of Massachusetts, Amherst MA; United States of America.\\
% ...
$^{125}$Homer L. Dodge Department of Physics and Astronomy, University of Oklahoma, Norman OK; United States of America.\\
% ...
$^{129}$LAL, Universit\'e Paris-Sud, CNRS/IN2P3, Universit\'e Paris-Saclay, Orsay; France.\\
% ...
$^{153}$Department of Physics and Astronomy, University of Sussex, Brighton; United Kingdom.\\
% ...
$^{164}$Department of Physics, University of Toronto, Toronto ON; Canada.\\
% ...
$^{a}$ Also at Borough of Manhattan Community College, City University of New York, NY; United States of America.\\
$^{b}$ Also at California State University, East Bay; United States of America.\\
$^{c}$ Also at Centre for High Performance Computing, CSIR Campus, Rosebank, Cape Town; South Africa.\\
$^{d}$ Also at CERN, Geneva; Switzerland.\\
$^{e}$ Also at CPPM, Aix-Marseille Universit\'e, CNRS/IN2P3, Marseille; France.\\
% ...
$^{*}$ Deceased

\end{flushleft}


\end{document}